\documentclass[fleqn,usenatbib]{mnras}
\usepackage{newtxtext,newtxmath}
\usepackage[dvipsnames]{xcolor}
\usepackage[T1]{fontenc}
\DeclareRobustCommand{\VAN}[3]{#2}
\let\VANthebibliography\thebibliography
\def\thebibliography{\DeclareRobustCommand{\VAN}[3]{##3}\VANthebibliography}

\usepackage{graphicx}	
\usepackage{amsmath}	
\usepackage{booktabs}
\usepackage{hyperref}
\usepackage{subcaption}
\usepackage{anyfontsize}

\title[SN 2023xwi: Early \lbrack Ca II\rbrack\, and \lbrack O I\rbrack]{\centering SN 2023xwi: Forbidden line emission in the peak spectrum of a Ca-strong transient}

\author[C.-G. Touchard-Paxton et al.]{C.-G. Touchard-Paxton$^{1,2}$\thanks{E-mail: toucharc@tcd.ie},
C. Frohmaier$^{1}$\thanks{E-mail: c.frohmaier@soton.ac.uk},
M. Pursiainen$^{3}$, 
M. Sullivan$^{1}$,
A. Polin$^{4}$,
G. Dimitriadis$^{5}$,
\newauthor
L. Galbany$^{6,7}$,
T. L. Killestein$^{8,3}$ ,
A. Kumar$^{3,9}$,
J. Lyman$^{3}$
\\
$^{1}$School of Physics and Astronomy, University of Southampton, University Road, Southampton, SO17 1BJ, UK.\\
$^{2}$School of Physics, Trinity College Dublin, the University of Dublin, College Green, Dublin 2, Ireland.\\ 
$^{3}$Department of Physics, University of Warwick, Gibbet Hill Road, Coventry, CV4 7AL, UK.\\
$^{4}$Department of Physics and Astronomy, Purdue University, 525 Northwestern Avenue, West Lafayette, IN 47907, USA.\\
$^{5}$Physics Department, Lancaster University, Lancaster, LA1 4YB, UK\\
$^{6}$Institute of Space Sciences (ICE-CSIC), Campus UAB, Carrer de Can Magrans, s/n, E-08193 Barcelona, Spain.\\
$^{7}$Institut d'Estudis Espacials de Catalunya (IEEC), 08860 Castelldefels (Barcelona), Spain.\\
$^{8}$Department of Physics \& Astronomy, University of Turku, Vesilinnantie 5, Turku, FI-20014, Finland.\\
$^{9}$Department of Physics, Royal Holloway - University of London, Egham Hill, Egham, TW20 0EX, UK.
}

\date{Accepted 2024 December 27. Received 2024 December 19; in original form 2024 September 27}

\pubyear{2024}

\begin{document}
\label{firstpage}
\pagerange{\pageref{firstpage}--\pageref{lastpage}}
\maketitle


\begin{abstract}
We present an extensive optical photometric and spectroscopic investigation into the calcium-rich supernova (SN) – SN2023xwi. Observations from a variety of ground-based telescopes follow the SN from 8 days pre-peak brightness to 87 days post-peak, covering both early-time (photospheric) and late-time (nebular) phases of the supernova. Objects of this class are characterised by nebular  spectra that are dominated by [Ca II] \(\lambda\lambda\) 7291, 7324 emission. SN 2023xwi displays a unique peculiarity in that its forbidden [Ca II] feature is visible in its peak photospheric spectrum – far earlier than expected in current models. This is one of the strongest and earliest detections of this feature in Ca-rich SNe in photospheric-phase spectra. We investigate the velocity evolution of this spectral feature and show that it cannot be easily explained by conventional progenitor systems. From our observations, we propose a SN progenitor embedded in an environment polluted by a recurrent He-nova AM CVn system.
\end{abstract}



\section{Introduction}

Ca-rich supernovae (SNe) are characterised by their strong forbidden [Ca II] emission in nebular-phase spectra, relative to the emission of [O I]. Although modelling of these SNe suggests that they do not produce a higher abundance of calcium compared to oxygen, they are noted for their prominent [Ca II] emission \citep{milisavljevic2017iptf15eqv,jacobson2022circumstellar}. Thus, while some have opted to call them “Ca-strong transients” (CaSTs; \cite{shen2019progenitors}), we refer to them as “calcium-rich transients” to reflect this characteristic. For Ca-rich SNe that occur in remote locations, the strong presence of [Ca II] indicates that the progenitors of these objects are sub-Chandrasekhar mass white dwarfs. The ejecta of more massive progenitors would continue to burn into heavier elements than Ca, resulting in spectra that would no longer be dominated by [Ca II] emission \citep{polin2021nebular}.

In addition to strong [Ca II] emission in their nebular-phase spectra, Ca-rich SNe share a unique set of behaviours: absolute magnitudes between that of novae and SNe, fast photometric evolution with rise times of < 15 days, photospheric velocities between 6,000 and 11,000  km s\textsuperscript{-1}, and rapid evolution to their nebular phase \citep{kasliwal2012calcium}. These transients are often found far from their host galaxies with no underlying star formation, indicating that they originate from old stellar populations \citep{lyman2014progenitors, lunnan2017two}. Ca-rich SNe typically have peak absolute magnitudes between -14 and -16.5 mag \citep{perets2010faint, kasliwal2012calcium, zenati2023origins}. This makes these objects far fainter than SNe Ia, which have peak absolute magnitudes of -19.5 mag \citep{branch1998type}. The combination of these features makes Ca-rich SNe particularly difficult to detect, so few have been observed, despite rate estimates suggesting they occur between 33 and 94\% as frequently as SNe Ia \citep{frohmaier2018volumetric}. Despite the challenges, investigating Ca-rich SNe remains imperative. These events not only challenge our understanding of thermonuclear SNe, but also offer insight into other areas of astrophysics. Ca-rich SNe may, for example, provide enough Ca to explain the observed excess of Ca in the intracluster medium (ICM) -- a long-standing mystery of the ICM \citep{mulchaey2013calcium}.

Historically, our knowledge of these exotic SNe has been limited by the small number of events we have been able to observe. The upcoming era of high cadence all-sky surveys such as the Vera C. Rubin Observatory's Legacy Survey of Space and Time (LSST), however, will significantly expand the available data set \citep{arendse2023detecting}. This will, in turn, allow us to gain a better understanding of the progenitor systems responsible for Ca-rich SNe.

In this paper, we present the transient SN 2023xwi which displays all of the characteristic behaviours of Ca-rich SNe. It also, however, has peculiar spectroscopic behaviour that defies our current understanding of objects of this class. From the earliest spectrum at +1d from peak, there is a distinct nebular forbidden [Ca II] feature. No explosion mechanisms of Ca-rich SNe allow for the photospheric phase ejecta to be at the low densities required to observe any nebular features \citep{dessart2015one}. The unique behaviour of SN 2023xwi therefore acts as a case study through which we can re-evaluate our understanding of the progenitor systems of Ca-rich SNe.

This paper presents a full spectroscopic and photometric analysis of SN 2023xwi. In Section \ref{observation} we present the discovery of SN 2023xwi and outline the photometric and spectroscopic observations. We determine physical parameters of the progenitor white dwarf from the photometric data in Section \ref{photo_analysis}. Section \ref{spec_analysis} covers the investigation of the evolution of the spectra, with a specific focus on the forbidden [Ca II] and [O I] features in Section \ref{early_neb}. In Section \ref{models} we discuss the pitfalls of current models, and present a plausible progenitor scenario that may explain the observed behaviour of SN 2023xwi. Finally, in Section \ref{summary} we give an overview of the work on SN 2023xwi and suggest future observations that could test the validity of the proposed progenitor system. 

Calculations in this paper assume a WMAP9 flat \(\Lambda\)CDM
cosmology with H\textsubscript{0} = 70 km s\textsuperscript{-1} Mpc\textsuperscript{-1} and \(\Omega\)\textsubscript{M} = 0.3. Throughout this paper, the phase is mentioned with respect to c-band peak brightness.


\section{Discovery and observations}
\label{observation}

\subsection{Discovery and classification}

\begin{figure}
	\includegraphics[width=\columnwidth]{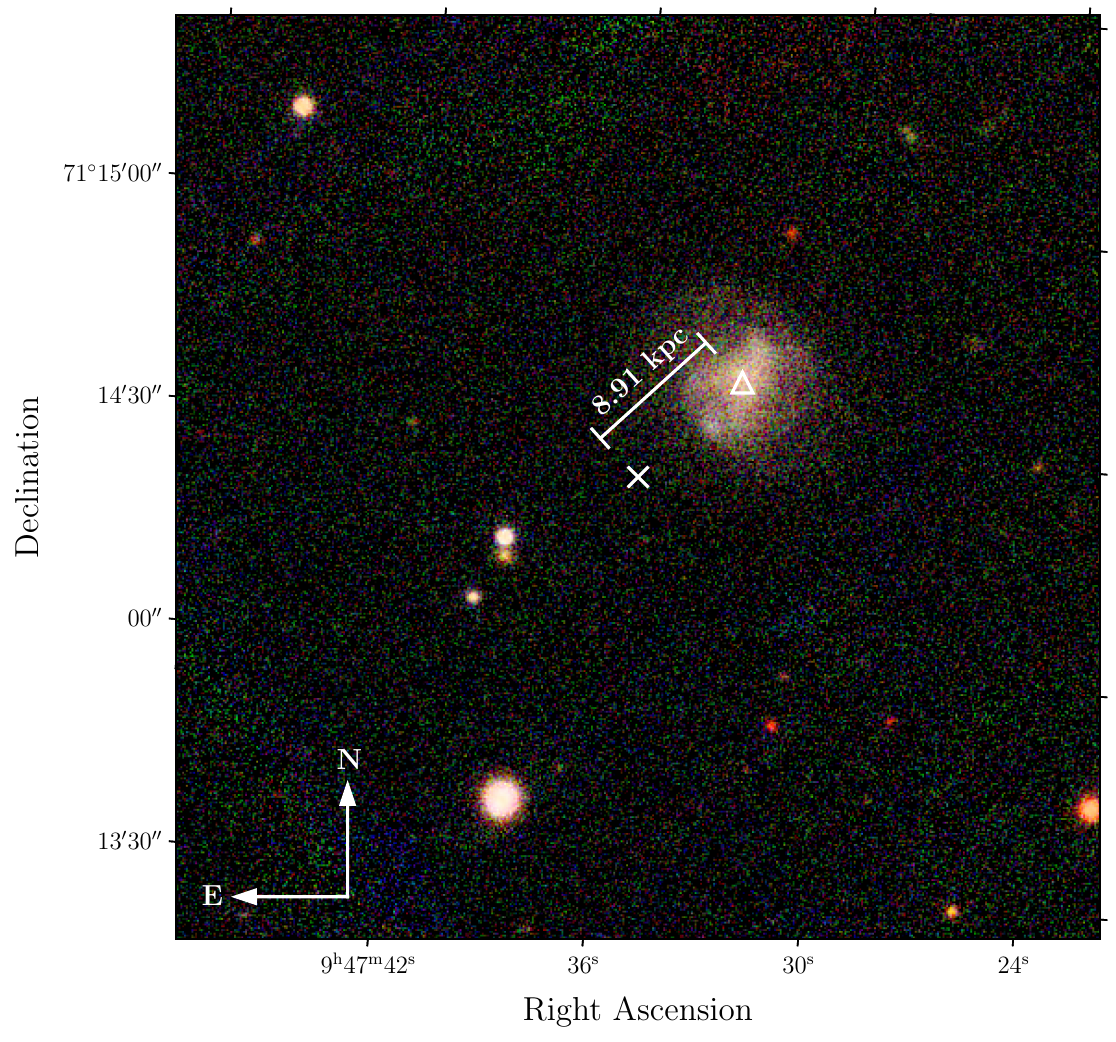}
    \caption{Colour composite image of the field of SN 2023xwi based on Pan-STARRS \emph{g}-, \emph{r}- and \emph{i}-band imaging data. SN 2023xwi (white cross) and its host PSO J146.8867+71.2435 (white triangle) have a separation of 8.91 \(\pm\) 0.17 kpc.}
    \label{fig:skymap}
\end{figure}

SN 2023xwi was first detected by the MASTER network on 15\textsuperscript{th} November 2023 (MJD 60263.9) with an internal name of MASTER OT J094735.55+711423.6 and was reported to the Transient Name Server (TNS) on 16\textsuperscript{th} November 2023 \citep{gress2023master}. Its apparent magnitude was measured to be \(\sim\)18.5 mag in open filter at coordinates \(\alpha\) = 09\textsuperscript{h}47\textsuperscript{m}35.540\textsuperscript{s} \(\delta\) = +71\textsuperscript{o}14'24.40", with a last non-detection date of 21\textsuperscript{st} November 2022. This initial detection was followed up with spectral analysis by Padova-Asiago using the Asiago Faint Object Spectrograph and Camera (AFOSC) on the Copernico 1.82m telescope on 18\textsuperscript{th} November 2023. The spectrum of SN 2023xwi closely matched that of SN 2016hgs, contributing to its classification of a SN Ib-Ca-rich with a redshift of z = 0.01. SN 2023xwi was also detected by Asteroid Terrestrial-impact Last Alert System (ATLAS; see \cite{tonry2018atlas}; ATLAS23vxx), the Gravitational wave Optical Transient Observatory (GOTO; see \cite{steeghs2022gravitational}; GOTO23bld), the Zwicky Transient Facility (ZTF; ZTF23abrgldj), and Gaia (Gaia23dpu). Fig. \ref{fig:skymap} shows the pre-explosion colour composite image we compiled using \emph{g}-, \emph{r}-, and \emph{i}-band Panoramic Survey Telescope and Rapid Response System (Pan-STARRS; see \cite{chambers2016pan}) data of the field of SN 2023xwi. The object is offset 8.91 \(\pm\) 0.1 kpc from its host galaxy PSO J146.8867+71.2435. 

We obtained spectra of the host galaxy from the Alhambra Faint Object Spectrograph and Camera (ALFOSC), using the [O III] \(\lambda\lambda\) 5007 feature in the host galaxy Nordic Optical Telescope (NOT) spectrum, we confirmed the redshift to be z = 0.0112 \(\pm\) 0.0005, corresponding to a luminosity distance of 48.4 \(\pm\) 1.2 Mpc to the host of SN 2023xwi.


\subsection{Photometry}
\label{photometry}


\begin{figure*}
	\includegraphics[width=0.9\linewidth]{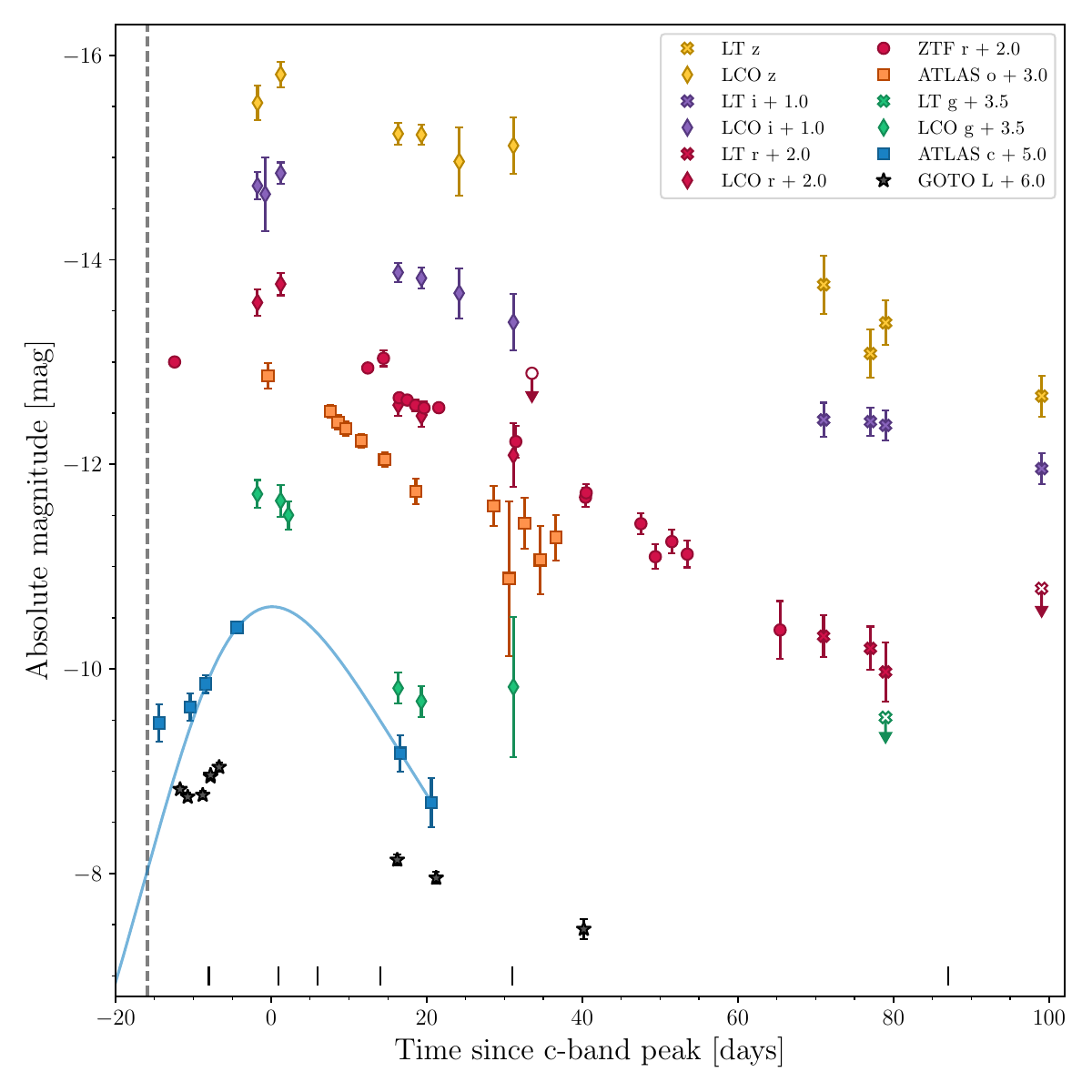}
    \caption{Multi-band rest frame light curves of SN 2023xwi showing the time evolution of its absolute magnitude in various filters (red: \emph{r}, orange: \emph{o}, yellow: \emph{i}, green: \emph{g}, blue: \emph{c}, purple: \emph{z}) relative to the \emph{c}-band peak. Data have been compiled from LCO (diamonds), LT (crosses), ATLAS (squares), ZTF (circles), and GOTO (stars), and have been offset in absolute magnitude by the value specified in the legend. Magnitudes were corrected for Galactic reddening (Section \ref{photometry}). Open tick marks represent upper limits of non-detections. A model of Bazin-like evolution has been overlaid onto the ATLAS \emph{c}-band data. Black vertical tick marks show the times at which spectra were observed. The dashed black vertical line shows the explosion epoch (see Section \ref{peak_time}).}
    \label{fig:light curve}
\end{figure*}

SN 2023xwi was imaged in \emph{g}-, \emph{r}-, \emph{i}-, \emph{z}-band at the Las Cumbres Observatory (LCO) and late-time follow-up imaging was sourced via Liverpool Telescope (LT). It was also imaged using ZTF in \emph{r}-band, ATLAS in \emph{c}- (4100 - 6600 \AA) and \emph{o}-band (5600 - 8100 \AA) \citep{tonry2018atlas}, and GOTO in \emph{L}-band (4000 - 7000 \AA) \citep{dyer2024gravitational}. LCO and LT data were reduced using tools available in \texttt{astropy} \citep{price2022astropy} and \texttt{photutils} \citep{larry_bradley_2024_10967176}. Final photometry was calibrated to the Pan-STARRS system. ZTF \emph{r}-band data, processed by the ZTF forced photometry pipeline \citep{masci2018zwicky}, have been included to cover the \(\sim\)40 day \emph{r}-band data gap between the last LCO \emph{r}-band detection and the first LT \emph{r}-band data point. ATLAS \emph{c}- and \emph{o}-band photometry, taken from the ATLAS forced-photometry server \citep{shingles2021release}, are the only observations of the source in epochs pre-, mid-, and post-peak. GOTO data, which were pre-reduced with forced photometry and their internal photometry pipeline, describe the pre-peak data with greater cadence than was possible with other telescopes. The compiled multi-band light curves from all observations between MJD 60260.5 and MJD 60373.9 from the above telescopes are shown in Fig. \ref{fig:light curve}. We corrected for foreground Galactic reddening using the \cite{schlafly2011measuring} recalibration of the \cite{schlegel1998maps} extinction maps. The Galactic reddening at the location of SN 2023xwi was E(B\(-\)V) \(= 0.056 \pm 0.003\) mag. We corrected this assuming a \cite{cardelli1989relationship} extinction law, with R\textsubscript{V} = 3.1. As SN 2023xwi is remote from its host, we did not correct for host extinction. We estimated the peak absolute magnitude of SN 2023xwi in \emph{r}-band to be -15.8 \(\pm\) 0.1 mag at MJD 60276.1. The non-detections, represented by open tick marks in Fig. \ref{fig:light curve}, are the upper 3-sigma limit of detections of SN 2023xwi. 

There is an early flux excess in the ATLAS \emph{c}-band and GOTO \emph{L}-band data before -10.7d from peak. This early flux excess has been noted before in a handful of Ca-rich SNe (e.g. \cite{jacobson2020sn, ertini2023sn}). Modelling of these double degenerate systems has shown that this can be reproduced by shock cooling emission or shock interaction with circumstellar material \citep{de2018iptf, jacobson2022circumstellar}, with some Ca-rich SNe (e.g. SN 2016hnk) likely being the result of thick He-shell detonation \citep{polin2019observational, jacobson2020hnk}. 


\subsection{Spectroscopy}

\begin{figure*}
	\includegraphics[width=0.9\linewidth]{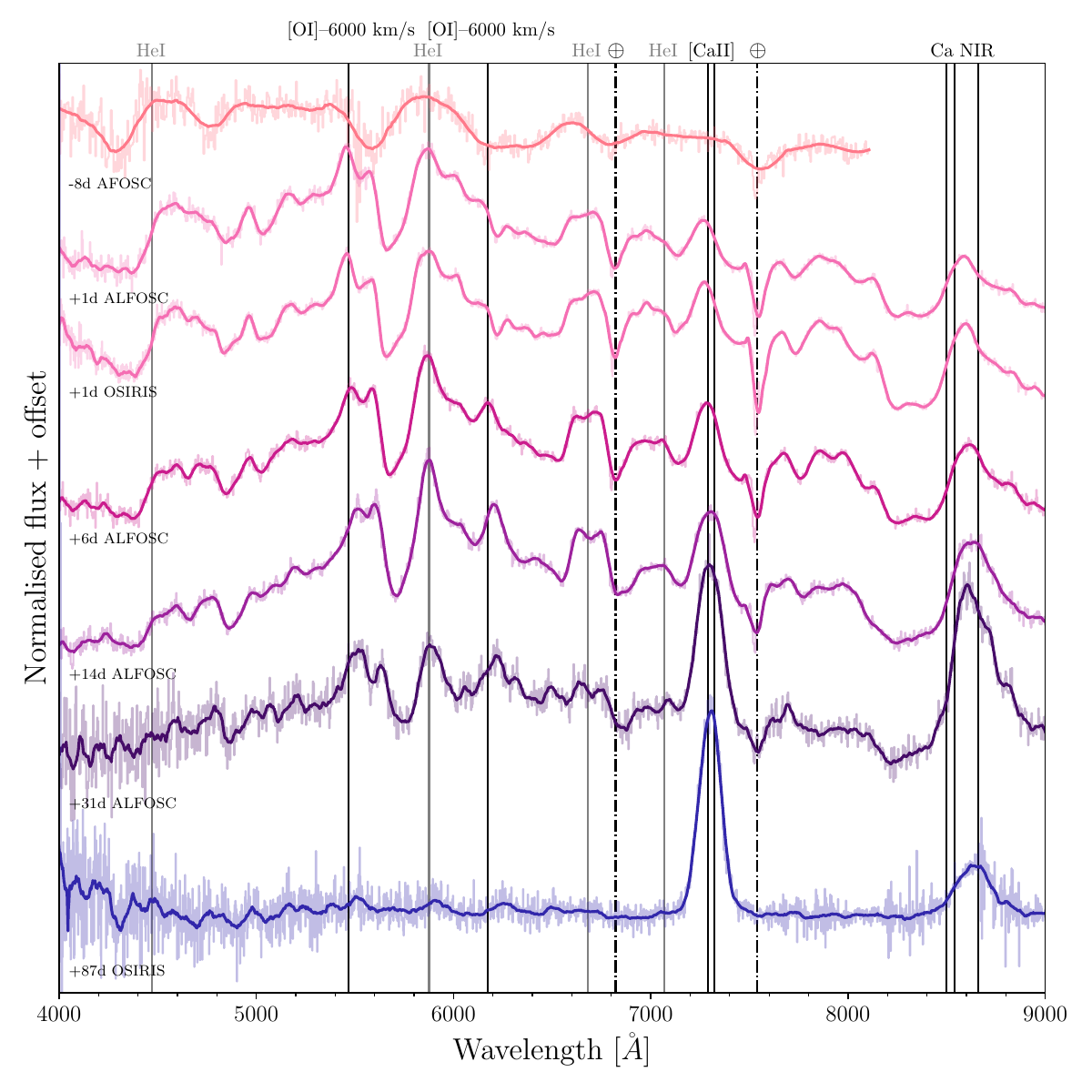}
    \caption{Spectral evolution of SN 2023xwi from -8d (top) to +87d (bottom) from \emph{c}-band peak obtained using AFOSC, ALFOSC, and OSIRIS. The raw data has been overlaid by a smoothed dataset using a Savitzky-Golay filter. The telluric absorption features are marked with a dot-dashed line. Each emission line of the nebular forbidden [Ca II] \(\lambda\lambda\) 7291, 7324 feature and the Ca II near-infrared (NIR) triplet \(\lambda\lambda\) 8498, 8542, 8662 are marked with black vertical lines. The wavelengths of the forbidden [O I] \(\lambda\lambda\) 5577, 6300 emission lines -- offset by 6000 km/s -- are marked with black vertical lines. The strong He I \(\lambda\lambda\) 4472, 5876, 6680, 7068 emission lines are marked with grey vertical lines. Phases shown are with respect to c-band peak brightness.}
    \label{fig:spec_ev}
\end{figure*}

The first spectrum of SN 2023xwi was taken at -8 d from peak with AFOSC and its similarity to SN 2016hgs at 6d post-peak led to its formal classification as Type Ib/c-Ca-rich \citep{de2018iptf}. Further follow-up spectra were observed using ALFOSC at the NOT and the Optical System for Imaging and low Resolution Integrated Spectroscopy (OSIRIS) at the Gran Telescopio CANARIAS (GTC) between MJD 60266.5 and MJD 60362.1 (+1 and 87 days post-peak, respectively). The NOT spectra were reduced with the PyNOT-redux reduction pipeline. Fig. \ref{fig:spec_ev} shows all spectra, overlaid with their smoothed data using a Savitzgy-Golay filter -- a function that models each successive group of small sets of the overall dataset as a low-order polynomial to increase signal precision with minimal distortion effects to its overall shape \citep{orfanidis1995introduction}. The telluric absorption features at \(\lambda\lambda\) 7615, 6890 are marked with a dot-dashed line on this figure.


\section{Photometric analysis}
\label{photo_analysis}

The light curve evolution of thermonuclear SNe is primarily driven by the mass of radioactive \textsuperscript{56}Ni synthesised in the explosion. This determines the peak brightness of a SN, its rise time, and influences the behaviour of its photosphere \citep{bora2022initial}. Investigating these characteristics of SN 2023xwi from its light curve (Fig. \ref{fig:light curve}), and inferring the mass of synthesised \textsuperscript{56}Ni, allows us to place further constraints on the progenitor systems of Ca-rich SNe. Any deviation of the photometric behaviour of SN 2023xwi from that of other members of its class may indicate abnormal amounts of synthesised \textsuperscript{56}Ni. This, in turn, could indicate that SN 2023xwi is the result of a particularly unique explosion mechanism or progenitor system.

The multi-band light curves of SN 2023xwi (Fig. \ref{fig:light curve}) display the expected behaviour of Type Ib/c Ca-rich SNe. Estimating the peak magnitude from the \emph{r}-band data gives a value of -15.8 \(\pm\) 0.1 mag at MJD 60276.1, with typical values of the class around -15.5 \citep{zenati2023origins}. The early flux excess, while not present in the light curves of all Ca-rich SNe, has been previously observed in a number of objects of this class (see Section \ref{photometry}).

\subsection{Time of peak brightness}
\label{peak_time}

We used the methods of \cite{bazin2009core} to determine the peak time of SN 2023xwi. As ATLAS \emph{c}-band was the only non-broadband filter covering the event before, during, and after peak, it was modeled with a Bazin-like evolution \citep{bazin2009core} of the form
\begin{equation}
   f(t) = A \frac{e^{-(t - t_0)/\tau _{\textrm{fall}}}}{1 + e^{-(t - t_0)/\tau _{\textrm{rise}}}} \textrm{,}
\end{equation}
with time t, and fit parameters A, \(\tau _{\textrm{rise}}\), \(\tau _{\textrm{fall}}\), t\textsubscript{0}. This model does not directly represent any physical processes occurring in the explosion, but instead describes a smoothly-evolving light curve with fit parameters that give insight into the general time evolution of the source. The normalisation of the magnitude is set by A, the rise and fall time variations are characterised by \(\tau _{\textrm{rise}}\) and \(\tau _{\textrm{fall}}\) respectively, and t\textsubscript{0} can be used to determine the date of peak, \(t_{\textrm{peak}}\), by
\begin{equation}
   t_{\textrm{peak}} = t_{0} + \tau_{\textrm{rise}}\ln\left(\frac{\tau_{\textrm{fall}}}{\tau_{\textrm{rise}}}-1\right) \textrm{.}
\end{equation}
Fitting this model to the ATLAS \emph{c}-band data gave a \emph{c}-band peak of 26\textsuperscript{th} November 2023 (MJD 60274.9 \(\pm\) 0.1). Tracing the ATLAS \emph{c}-band model back to the limit of non-detection gives an explosion date of 11\textsuperscript{th} November 2023 (MJD 60259.1 \(\pm\) 0.1), a result which is consistent with the data presented in the light curve. This gives a rise time of 15.9 \(\pm\) 0.2 days. Ca-rich SNe are expected to have rise times around 15 days \citep{kasliwal2012calcium}, confirming SN 2023xwi to have a rise time that is typical of objects of this class. 


\subsection{Properties of ejecta}
\label{ejecta_properties}


\subsubsection{\texttt{SuperBoL}}
\label{superbol}

To investigate the properties of the ejecta of SN 2023xwi, we synthesised a pseudo-bolometric light curve using \texttt{SuperBoL} -- an open-source Python package that calculates the bolometric light curves of SNe using observed photometric data \citep{nicholl2018superbol}. The quasi-bolometric (or pseudo-bolometric) method within \texttt{SuperBoL} combines time-series multi-colour photometric data across the broadest possible wavelength range. Ideally, data would be available across all wavebands -- ultraviolet (UV) to infrared (IR) -- but this is often not possible. Using its inbuilt capacity to approximate black body flux outside of the observed wavelengths, \texttt{SuperBoL} can create a pseudo-bolometric light curve from a narrow range of filters. 

Running the code allows for black body fitting across all wavelengths, cosmological distance corrections, extinction corrections, and integration of the synthesised pseudo-bolometric light curve. By integrating the flux of the input SN, \texttt{SuperBoL} determines the luminosity at each of the key epochs. This package is based on the assumption that the photosphere remains spherical throughout its evolution. As SNe are neither intrinsically spherical nor perfect black bodies, the calculated black body values therefore should only be treated as approximate estimates which describe the time evolution behaviour of the black-body parameters. We generated the rest-frame quasi-bolometric light curve of SN 2023xwi, covering the g, r, i, z, o, and c bands. Constant colours were assumed to compute the magnitudes at common epochs. To account for the expected flux contribution from the UV and NIR regions, we extrapolated the SED -- using the black body model -- by integrating over the observed fluxes and obtained the full bolometric light curve.

The peak luminosity of SN 2023xwi, determined from this pseudo-bolometric light curve, is 2.2 \(\pm\) 0.3 \(\times\) 10\textsuperscript{41} erg s\textsuperscript{-1}. \cite{kasliwal2012calcium} report a similar peak luminosity of 4.6 \(\times\) 10\textsuperscript{41} erg s\textsuperscript{-1} for PTF 10iuv, and \cite{jacobson2022circumstellar} found the peak luminosities of SN 2021gno and SN 2021inl to be 4.12 \(\times 10^{41}\) and 2.37 \(\times 10^{41}\) erg s\textsuperscript{-1}, respectively. This confirms that SN 2023xwi has a similar peak luminosity to that of other members of its class.

The black body radius expands from 0.6 \(\pm\) 0.1 \(\times\) 10\textsuperscript{15} cm at peak to 1.2 \(\pm\) 0.3 \(\times\) 10\textsuperscript{15} cm at 14d post-peak, at which point it begins to recede. As nebular features are only expected to be present in the spectra of SNe once the photosphere recedes into inner layers of the explosion, it can be approximated that nebular features should not be visible in SN 2023xwi until well after this time of 14d post-peak. This value is consistent with other Ca-rich SNe where some tentative signatures of nebular features are seen from \(\sim\)+14d \citep{de2020zwicky, jacobson2022circumstellar}. As we discuss in Section \ref{spec_analysis}, however, this is not the case for SN 2023xwi which shows nebular features earlier than this estimate.

\subsubsection{\textsuperscript{56}Ni mass}

As the peak brightness of a thermonuclear SN is directly related to the amount of \textsuperscript{56}Ni synthesised in the explosion of the CO white dwarf \citep{arnett1982type}, determining the synthesised mass of \textsuperscript{56}Ni in a SN allows us to place further constraints on the responsible explosion mechanisms. Double detonation models for Ca-rich SNe predict \textsuperscript{56}Ni masses between 0.01 and 0.05 M\textsubscript{\(\odot\)} \citep[e.g.][]{zenati2023origins} -- any significant deviations from this value would require the modification of current explosion models.

As SN 2023xwi was initially classified as a Type Ib/c-Ca-rich, we used the methods of \cite{khatami2019physics} to determine the mass of nickel synthesised in the explosion and the total ejecta mass. The peak time-luminosity relation in Type Ib/c SNe is given by
\begin{equation}
\begin{split}
    L_{\textrm{peak}} = \frac{2 \epsilon_{\textrm{Ni}} M_{\textrm{Ni}}}{{\beta}^2 {\tau_{\textrm{Ni}}}^2} [0.83(1 - \beta \tau_{\textrm{Ni}})e^{-\beta \tau_{\textrm{Ni}}} \\ + 26.56(1 - (1 + \beta \tau_{\textrm{Co}})e^{-\beta \tau_{\textrm{Co}}})] \textrm{,}
\end{split}  
\end{equation}
with peak luminosity \(L_{\textrm{peak}}\), specific heating rate of Ni-decay \(\epsilon_{\textrm{Ni}}\), mass of nickel synthesised in the explosion \(M_{\textrm{Ni}}\), the opacity and concentration-dependent factor \(\beta\), and decay-time parameters \(\tau_{\textrm{Ni}}\) and \(\tau_{\textrm{Co}}\). \(\tau_{\textrm{Ni}}\) and \(\tau_{\textrm{Co}}\) are related to the decay timescales of nickel and cobalt (\(t_{\textrm{Ni}}\) = 8.8 days and \(t_{\textrm{Co}}\) = 111.3 days) by \(\tau_{\textrm{Ni}} = t_{\textrm{peak}} / t_{\textrm{Ni}}\) and \(\tau_{\textrm{Co}} = t_{\textrm{peak}} / t_{\textrm{Co}}\) where \(t_{\textrm{peak}}\) is the time between first light and peak of the SN. To calculate the mass of nickel synthesised by SN 2023xwi, we used \(\epsilon_{\textrm{Ni}}\) = 3.9 \(\times\) 10\textsuperscript{10} erg g\textsuperscript{-1} s\textsuperscript{-1} and \(\beta\) = 9/8 \citep[see Table 2 of][]{khatami2019physics}.  Using \(L_{\textrm{peak}}\) = 2.2 \(\pm\) 0.3\(\times 10^{41}\) erg s\textsuperscript{-1}, given by the peak luminosity of the pseudo-bolometric light curve, and rise time t\textsubscript{rise} = 15.9 days from the Bazin-like model yields a nickel mass of 0.02 \(\pm\) 0.01 M\textsubscript{\(\odot\)}. This is consistent with the expected range of 0.01 to 0.05 M\textsubscript{\(\odot\)} \citep{ertini2023sn, zenati2023origins}.

\subsubsection{Ejecta mass}
In addition to the \textsuperscript{56}Ni mass discussed above, estimating the ejecta mass of thermonuclear SNe allows us to further constrain current models of their explosion mechanisms \citep{zenati2023origins}. 

Following the methods of \cite{khatami2019physics}, the total ejecta mass \(M_{\textrm{ej}}\) can be determined from the diffusion timescale \(t_{\textrm{d}}\) using
\begin{equation}
   t_{\textrm{d}} = \sqrt{\frac{\kappa M_{\textrm{ej}}}{v_{\textrm{ej}}c}} \textrm{,}
\end{equation}
with electron scattering opacity \(\kappa\), ejecta velocity \(v_{\textrm{ej}}\), and speed of light \(c\). We used a value of \(t_{\textrm{d}}\) = 32.0 \(\pm\) 0.1 days, given by the equation 
\begin{equation}
   \frac{t_{\textrm{peak}}}{t_{\textrm{d}}} = 0.11 \ln{\left(1 + \frac{9\:t_{\textrm{Ni}}}{t}\right)} - 0.36 \textrm{,}
\end{equation} 
with \(t_{\textrm{peak}}\) calculated using the method discussed above. 

Using the assumption that the ejecta is dominated by iron-group elements, we took \(\kappa\) to be 0.1 cm\textsuperscript{2} g\textsuperscript{-1} \citep{piro2013can, dimitriadis2023sn}. The ejecta velocity can be determined from the absorption components of the He I \(\lambda\lambda\) 5876 and Ca II NIR features in photospheric spectra. At +1d, the velocities of these features were \(v_{\textrm{He I}}\) = 10,900 \(\pm\) 100 km s\textsuperscript{-1} and \(v_{\textrm{NIR}}\) = 8,100 \(\pm\) 100 km s\textsuperscript{-1}, yielding total ejecta masses of \(M_\textrm{ej}\) = 1.2 \(\pm\) 0.2 M\textsubscript{\(\odot\)} and \(M_\textrm{ej}\) = 0.9 \(\pm\) 0.2 M\textsubscript{\(\odot\)} respectively. It should be noted that these calculated values are overestimates of the true values of the mass of the ejecta as the methodology relies on the assumption that the SN is spherical. As SNe are intrinsically `messy' events, the calculated ejecta mass value of M\textsubscript{ej} = 1.2 M\textsubscript{\(\odot\)} can serve only as an upper limit \citep{maeda2010asymmetric}.


\section{Spectroscopic analysis}
\label{spec_analysis}

Fig. \ref{fig:spec_ev} shows the full spectral evolution of SN 2023xwi between \(-8\)d and +87d of the \emph{c}-band peak. The photospheric spectra (i.e. up to +14 d) show a distinct He I \(\lambda\lambda\) 5876 feature. By modelling the absorption feature as a Gaussian profile, we measured the associated velocity at peak to be 10,910 \(\pm\) 100 km s\textsuperscript{-1} -- a value consistent with the population of Ca-Ib/c events presented in \cite{de2020zwicky}. The nebular spectra show prominent forbidden [Ca II] and Ca II (NIR) features -- the signature of objects of this class. 

SN 2023xwi notably differs from other Ca-Ib/c and Ca-rich SNe in general due to two key features in its spectra: the distinct presence of [Ca II] emission at peak, and the presence of forbidden [O I] emission from 6d post-peak. We discuss these features in Section \ref{early_neb}.


\subsection{Spectral decomposition}

\subsubsection{\texttt{SYNAPPS}}

To gain further insight into the chemical composition of the ejecta of SN 2023xwi, we modelled each photospheric phase spectrum using \texttt{SYNAPPS} -- a highly parametrised synthetic spectrum-fitting software \citep{thomas2011synapps}. \texttt{SYNAPPS} is an automated implementation of \texttt{SYN++} -- a local thermodynamic equilibrium (LTE) spectral synthesis code based on the following assumptions: spherical symmetry, homologous expansion, a sharp photosphere that emits a continuous black body spectrum, and line formation by resonance scattering -- a quantum mechanical effect that arises from the interaction between photons and atoms / molecules \citep{fisher1997evidence, parrent2010synow, kumar2021sn}. The resonance scattering is treated in the Sobolev approximation --  an approximation of the solution of the radiative transfer equation, assuming that local velocity gradients are negligible compared to the overall velocity gradient \citep{jeffery1990analysis}. The primary aim of \texttt{SYN++} is to aid line identification and estimations of photospheric velocity. \texttt{SYNAPPS}, the automated version of this, is based on the same assumptions. These approximations are suited to high-density environments, making \texttt{SYNAPPS} a suitable tool for the modeling of photospheric spectra \citep{thomas2011synapps}. As our focus was with the line identification of photospheric spectra, it was not necessary to use more complex and computationally expensive non-LTE codes.

Spectra of Ca-rich SNe have also been modelled using non-LTE codes \citep[e.g.][]{dessart2015one, zenati2023origins}, that account for radiative energy leakage from the rapidly expanding, low density, ejecta of the SN. Intriguingly, \cite{dessart2015one} show a strong contribution from Ti II is possible around peak in the region we identify as [Ca II] (Section \ref{CaII}) and even a blended contribution from Sc II and Ti II near where we identify [O I] (Section \ref{OI}). When compared to literature examples of Ca-rich SNe, however, they find that these models are redder due to line blanketing from the large amount of Ti II. Given that SN 2023xwi is more similar to the literature sample of Ca-rich SNe than to these models, we proceed our analysis under the assumption that strong [Ca II] and [O I] are present in our photospheric phase spectra. We do encourage detailed, non-LTE modelling of SN 2023xwi and other Ca-rich SNe to conclusively identify and better understand the origin of these features.

\subsubsection{Photospheric spectra}
\label{6dspec}

Various combinations of component ions were tested for each spectrum. Each \texttt{SYNAPPS} fit was optimised using the AIC (Akaike Information Criterion) and BIC (Bayesian Information Criterion) values for each iteration of ions tested. AIC and BIC are defined as
\begin{equation}
   \textrm{AIC} = -2 \ln{(\textrm{L})} + 2k
\end{equation}
and
\begin{equation}
   \textrm{BIC} = -2 \ln{(\textrm{L})} + 2k \ln{(\textrm{N})} \textrm{,}
\end{equation}
with likelihood L, number of model parameters k, and number of data points N \citep{akaike1974new, stone1979comments}. For each fit, we calculated the AIC and BIC values using \texttt{RegscorePy} -- a Python package that aids model comparison  with the use of various regression models. AIC does not penalise as harshly as BIC, and can therefore favour more complex models provided the fit is significantly better. The favoured model using AIC therefore may be over-fitting the data, and as such may not be physically plausible. Conversely, BIC penalises harshly for additional parameters, and will favour the simplest appropriate model. This can result in a favoured model that is an over-simplification of the underlying physical mechanisms. Due to the intricacies of each criterion, it is best to optimise a fit with respect to both the AIC and BIC values to ensure the model is neither unnecessarily complicated nor overly-simplified. 

\begin{figure}
	\includegraphics[width=\columnwidth]{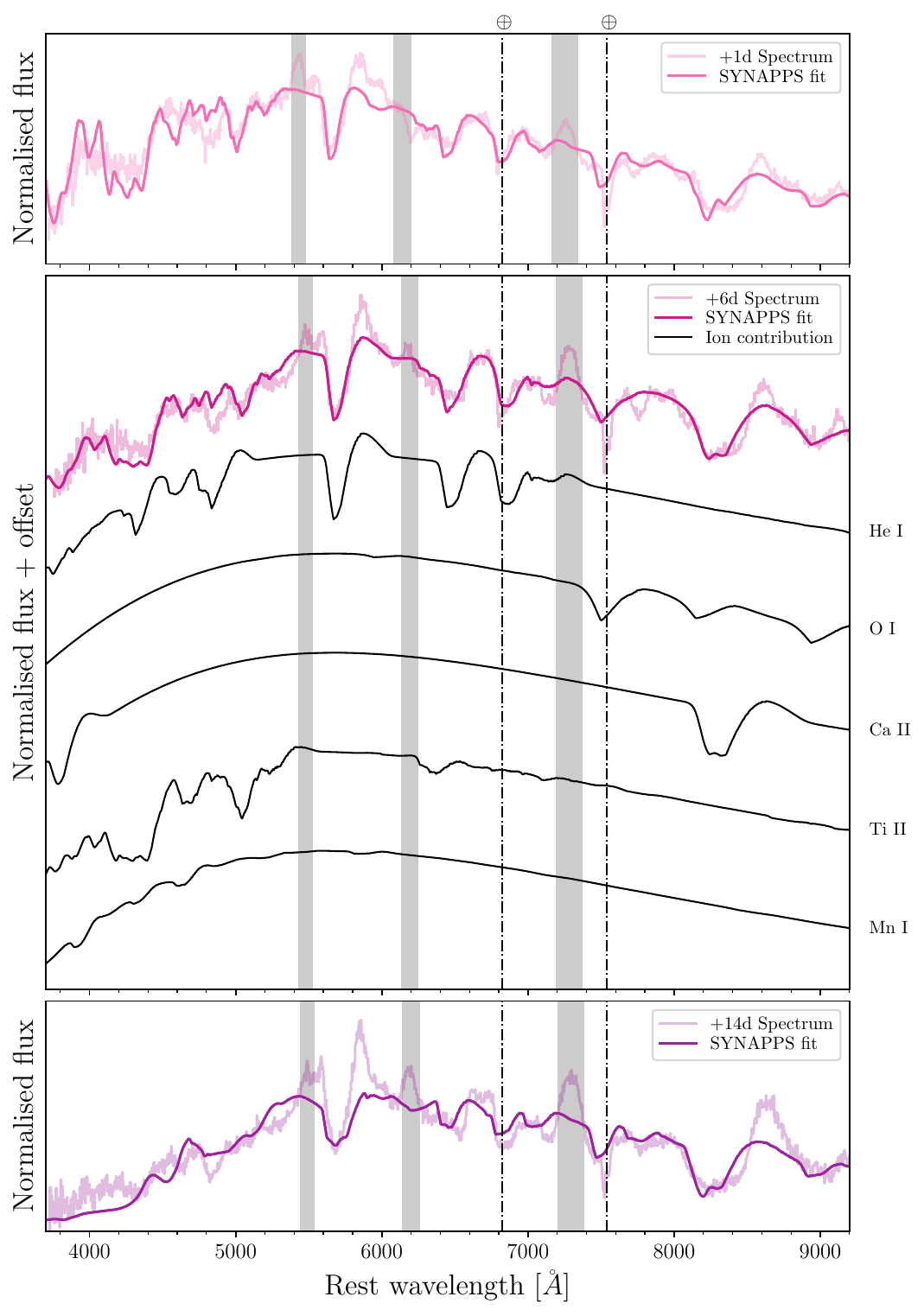}
    \caption{Top: \texttt{SYNAPPS} fit of the +1d spectrum of SN 2023xwi. Middle: spectral decomposition of the +6d spectrum. The data (light pink) have been overlaid with the \texttt{SYNAPPS} fit (dark pink). The [Ca II] feature at 7,308 {\AA}, along with the telluric absorption features (dot-dashed vertical lines) were masked in the fitting process. The contribution of each fitted ion to the overall spectrum fit is shown by black traces in order of ion mass: He I (top) to Mn I (bottom). The flux of each ion contribution has been normalised relative to the overall spectrum  fit. Bottom: \texttt{SYNAPPS} fit of the +14d spectrum. Dot-dashed lines show the telluric features which were masked from all \texttt{SYNAPPS} fits. The wavelengths that were additionally masked from the \texttt{SYNAPPS} fits in the `masked' section of analysis (see Table \ref{tab:aic_bic}) have been highlighted in grey. Phases shown are with respect to c-band peak brightness.}
    \label{fig:synapps_all}
\end{figure}

Fig. \ref{fig:synapps_all} shows the observed spectrum and the best \texttt{SYNAPPS} fit to each of SN 2023xwi's photospheric spectra. The ionic species included in the final \texttt{SYNAPPS} model for each spectrum were He I, O I, Ca II, Ti II, and Mn I. The middle panel shows the contribution of each ionic species -- extracted from the total fit using \texttt{SYN++} -- to the total spectrum of SN 2023xwi at +6d after \emph{c}-band peak. While the \texttt{SYNAPPS} model is an overall good fit, there are three spectral features it cannot model in each spectrum: the emission features at \(\sim\)5,500 {\AA}, \(\sim\)6,300 {\AA}, and \(\sim\)7,300 {\AA}. To ensure their presence would not distort the fit, these emission features were masked from the input spectrum. The start and end points of the masked sections were taken to be the local minima on either side of each emission feature. As discussed in Section \ref{superbol}, we would expect SN 2023xwi to start to transition to its nebular phase at +14d. Ca-rich objects rapidly transition to their nebular phase, at which point \texttt{SYNAPPS} struggles to model their spectra. As a result, the \texttt{SYNAPPS} fit at +14d deviates more from the observed spectrum than the fits from the earlier, more truly-photospheric, spectra.

\begin{table}
  \begin{center}
    \caption{AIC and BIC values for the \texttt{SYNAPPS} model fit to unmasked and masked +1d, +6d, and +14d input spectra of SN 2023xwi. `Unmasked' refers to the normalised spectra with only the telluric absorption features masked, and `masked' refers to additional masking of the emission features at \(\sim\)7,300, \(\sim\)6,300 {\AA}, and \(\sim\)5,500 {\AA}. A lower value of both AIC and BIC for the same input data implies a better fit. Phases mentioned are with respect to c-band peak brightness.}
    \label{tab:aic_bic}
    \begin{tabular}{l|cc|cc}
      \toprule
       & \multicolumn{2}{c}{\textbf{Unmasked}} & \multicolumn{2}{c}{\textbf{Masked}}\\
       \cmidrule(lr){2-3}\cmidrule(lr){4-5}
       & AIC & BIC & AIC & BIC\\
      \midrule
      +1d & -6510 & -6448 & -8753 & -8705\\
      +6d & -7129 & -6742 & -8918 & -8870\\
      +14d & -6696 & -6613 & -7906 & -7858\\
      \bottomrule
    \end{tabular}
  \end{center}
\end{table}

The AIC and BIC values before and after the peculiar emission features were masked in each spectrum are given in Table \ref{tab:aic_bic}. The `unmasked' values are the AIC and BIC values for the normalised spectra, and the `masked' values are those for the spectra where the emission features at \(\sim\)7,300, \(\sim\)6,200 {\AA}, and \(\sim\)5,500 {\AA} were masked in the \texttt{SYNAPPS} fit. The spectra at +31 and +87d have been omitted as they are too nebular for \texttt{SYNAPPS} fitting to be appropriate. The fit of the \texttt{SYNAPPS} models to the masked data is significantly improved from the unmasked data in each of the photospheric spectra. This indicates that the emission features are not expected for photospheric-phase spectra -- it is instead more likely that these are nebular phase emission features. 

As discussed in Section \ref{superbol}, we would not expect to see nebular features in the spectra of SN 2023xwi until well after +14d from peak. The presence of these three nebular-phase features so early in the spectra is not predicted by current progenitor systems of Ca-rich SNe as the ejecta at this phase would be too dense for the necessary forbidden transitions to occur. 

\section{Early nebular emission}
\label{early_neb}

The wavelengths of the peculiar emission features identified in Section \ref{6dspec} indicate that they are [Ca II] and [O I]. These are typically nebular-phase features -- they are caused by emission cooling of the centre of the SN and are visible after the photosphere has receded to reveal the core of the explosion. As was estimated in Section \ref{ejecta_properties}, the photosphere does not begin to recede until \(\sim\)14d post-peak, and thus we would not expect to observe nebular features until well after this time. In the nebular phase, the ejecta have reached low-enough ion number densities for these transitions to occur (\(\sim\) 10\textsuperscript{7} g cm\textsuperscript{-3} for [Ca II]). The presence of these features in the photospheric spectra, a phase where the ejecta densities are greater than this critical density, is therefore unexpected. To determine whether the presence of these typically nebular-phase features is compatible with our current understanding, we investigate the nature and evolution of these features below.

\subsection{Forbidden [Ca II] feature}
\label{CaII}

The spectral time series of SN 2023xwi (see Fig. \ref{fig:spec_ev}) shows clear presence of the [Ca II] \(\lambda\lambda\) 7291, 7324 emission feature from the peak-phase spectra out to late times. Peak [Ca II] emission has only previously been detected in one other Ca-rich SN -- SN 2019ehk \citep{jacobson2020sn}. In this section we present analysis of the line profiles and evolution of [Ca II] in SN 2023xwi. We also couple this with our unique detection of [O I] around peak (see Section \ref{OI}), to speculate on the physical origin of our forbidden emission and present a progenitor scenario in Section \ref{models}.

\subsubsection{[Ca II] velocity evolution}
\begin{figure*}
    \includegraphics[width=\linewidth]{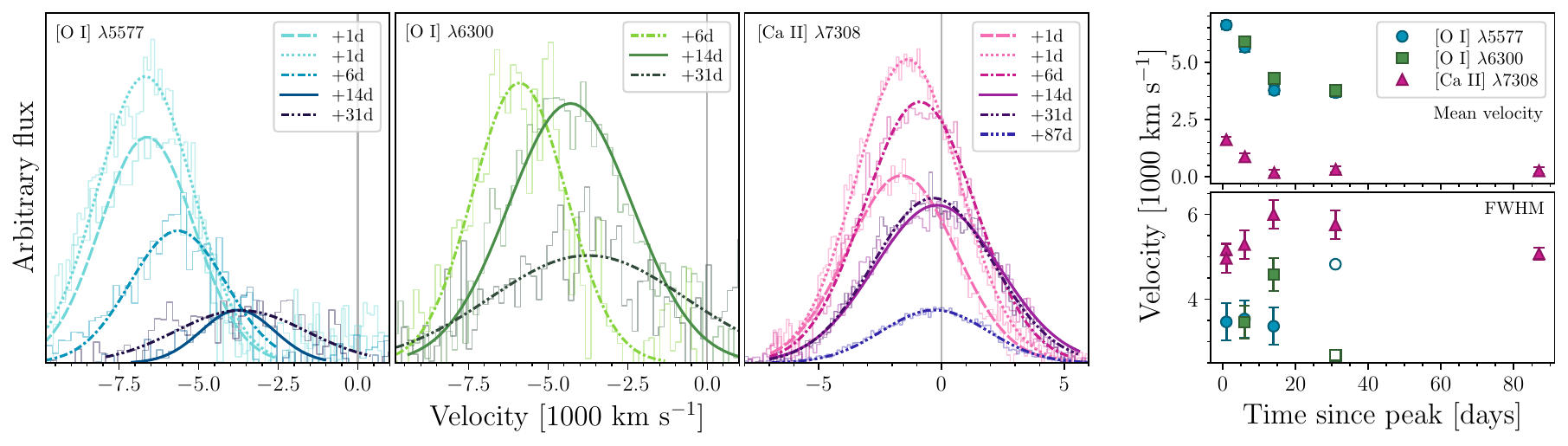}
    \caption{Left panel: Continuum-corrected velocity-space emission features for [O I] \(\lambda\lambda\)5577 (blue, left), 6300 (green, centre left), and [Ca II] \(\lambda\lambda\) 7308 (pink, centre right). Relative to the \emph{c}-band peak, dashed (ALFOSC) and dotted (OSIRIS) lines show the features at +1d, dot-dashed lines at +6d, solid lines at +14d, dot-dot-dashed lines at +31d, and dot-dot-dot-dashed lines at +87d. Right panel, top: the velocity evolution of the three features with the 1\(\sigma\) uncertainties associated with each value. Right panel, bottom: full width at half maximum (FWHM) evolution of the three features with the 1\(\sigma\) uncertainties associated with each value. Hollow tick marks show the estimated [O I] FWHM at +31d. Phases shown are with respect to c-band peak brightness.}
        \label{fig:velocity_ev}
\end{figure*}

Fig. \ref{fig:velocity_ev} shows the continuum-corrected velocity evolution and full width at half maximum (FWHM) evolution of the [Ca II] feature of SN 2023xwi in its spectra between 1d and 87d post-peak. We modelled the feature as a combination of the local pseudo-continuum and the profile of the emission lines. The local pseudo-continuum was approximated by a linear relationship between wavelength and observed flux between the two local minima of the emission feature. The emission profile was modelled as a singular Gaussian profile as the two component emission lines of this feature could not be resolved. Uncertainties were calculated following one of the standard routine for least squares minimisation within \texttt{LMFIT} \citep{newville2016lmfit}. The mean velocity associated with this feature is 1,600 \(\pm\) 100 km s\textsuperscript{-1} at peak. This is far less than the photospheric velocity of 10,900 \(\pm\) 100 km s\textsuperscript{-1}, and therefore the observed feature cannot be the result of Ca in the photosphere. The mean velocity of the feature decreased with time to 300 \(\pm\) 100 km s\textsuperscript{-1} at 87d post-peak, a value which is consistent with zero at a significance level of 2.6\(\sigma\). As the nebular spectra show the intrinsic properties of the core of the SN, we expect the velocity associated with the [Ca II] emission feature to be 0 km s\textsuperscript{-1}. This is not the case for the [Ca II] features at +1d and +6d of SN 2023xwi which have a velocity inconsistent with 0 to 6\(\sigma\). The FWHM remains constant across all phases at 5,300 \(\pm\) 200 km s\textsuperscript{-1} at a significance level of 2.3\(\sigma\). This implies the velocity dispersion of the emitting region remains constant and therefore there is no significant deceleration of the ejecta due to interaction with dense external material. The large difference between the mean [Ca II] velocity and the FWHM suggests the [Ca II] emission originates from a narrow region which is extended at a large radius from the core of the SN \citep{taubenberger2019sn}.

\cite{prentice2020rise} presented a similar early nebular feature of [Ca II] in SN 2019bkc. This feature was present in the spectra from 13d post-peak and had an associated velocity of 10,000 km s\textsuperscript{-1}. This source evolved very rapidly, with a rise time of \(\sim\)6 days, so the early presence of [Ca II] was concluded to be due to early evolution to the nebular phase, and not nebular emission within the photospheric-phase spectra as is seen in SN 2023xwi. There was no clear origin of the high velocity [Ca II] feature, but one suggestion was that the measured high velocities were the result of observing a `blob' of ejected material that was in front of the photosphere. In this model, the emission is explained as excitation from radioactive material in the `blob'. As the [Ca II] velocity in SN 2023xwi is lower than the photospheric velocity, it is unlikely that a `blob' was ejected from the SN as it is not moving at a great enough velocity to escape the photosphere, and as such would not be observable. Additionally, as the [Ca II] feature was not present in the spectra of SN 2019bkc until +13d, the `blob' model does not explain the presence of [Ca II] emission in the photospheric spectra of SN 2023xwi at +1d.

It could be that the photosphere overtook an external region of low-density Ca between +6 and +14d (velocities of 900 \(\pm\) 100 and 200 \(\pm\) 100 km s\textsuperscript{-1}, respectively) and this is why we observe two distinct velocities of \(\sim\)0 km s\textsuperscript{-1} (after +14d) and \(\sim\)1600 km s\textsuperscript{-1} (up to +6d). The higher velocity would therefore correspond to the low-density region, and the zero-velocity to the true nebular-phase feature of the SN itself. 

For the above to be plausible, an area of low-velocity low-density Ca would have to be in close-proximity to the progenitor system before the SN event. Ca is synthesised during the SN, so current models do not account for the presence of sufficient quantities of Ca in the system pre-explosion. We are therefore left with one key question: how does a region of low-density low-velocity Ca exist before the SN?

\subsubsection{Estimating synthesised Ca mass}

Following the methods of \cite{jacobson2022circumstellar}, the relationship between the mass of calcium synthesised in the explosion and the luminosity of the nebular [Ca II] feature is given by
\begin{equation}
    L_{\textrm{[Ca II]}} = V_{\textrm{ph}} n_{\textrm{[Ca II]}} A_{\textrm{[Ca II]}} h \nu_{\textrm{[Ca II]}} \left(\frac{10}{11}\right) e^{-19700/T} \textrm{,}
        \label{luminosity_ca}
\end{equation}
with luminosity of the [Ca II] feature L\textsubscript{[Ca II]}, volume of photosphere V\textsubscript{ph}, Einstein A coefficient A\textsubscript{[Ca II]} = 2.6 s\textsuperscript{-1}, number density of Ca II ions n\textsubscript{Ca II}, energy of a photon h$\nu$\textsubscript{[Ca II]}, temperature of the photosphere T, and statistical weighting given by the fraction (10/11). This relation is appropriate when the [Ca II] feature is in its high-density limits, i.e. ion number densities greater than 10\textsuperscript{7} g cm\textsuperscript{-3}. Using a range of temperatures of T = 5,000 to 10\textsuperscript{4} K, and [Ca II] feature maximum integrated luminosity L\textsubscript{[Ca II]} = \(2.7 \times 10^{39}\) erg s\textsuperscript{-1} yields values of Ca II ion number density n\textsubscript{Ca II} = \(2.3 \times 10^{7}\) and \(0.3 \times 10^{7}\) g cm\textsuperscript{-3}. As only the low temperature limit of 5,000 K yields an appropriate ion number density for use of Equation \ref{luminosity_ca}, we only use this temperature value in the further calculations.

Using the same photospheric volume as above to determine the number of Ca II ions synthesised gives an upper limit of synthesised calcium mass M\textsubscript{Ca} = \(9 \times 10^{-4} \) M\textsubscript{\(\odot\)} for 5,000 K. This is consistent with the values presented in \cite{jacobson2022circumstellar} which determined the synthesised calcium mass for the same temperature as M\textsubscript{Ca} = 9 \(\times 10^{-4}\) M\textsubscript{\(\odot\)} for SN 2021gno, and M\textsubscript{Ca} = 10 \(\times 10^{-4}\) M\textsubscript{\(\odot\)} for SN 2021inl. The consistency between these values confirms that a standard mass of Ca is synthesised in SN 2023xwi. It therefore rules out the possibility of the abnormally strong early forbidden [Ca II] emission feature being a byproduct of increased Ca synthesis. 

It should be noted that as the final spectrum at +87d is not fully nebular, the derived mass of calcium is likely to be lower than the true elemental mass in the explosion. Despite this, as the mass estimates for SN 2023xwi are consistent with those presented in \cite{jacobson2022circumstellar}, this method proves to be a useful metric in determining whether the mass of Ca synthesised in a Ca-rich SN is abnormal.


\subsection{Forbidden [O I] feature}
\label{OI}

As is the case for the [Ca II] feature discussed above, the [O I]\(\lambda\lambda\) 5,577, 6,300 features are forbidden transitions, and as such should only be present in the nebular spectra. The [O I] \(\lambda\lambda\) 6300 feature is present in all spectra from +6 to +31d, and is blended into other spectroscopic features in the +1d spectra. The [O I] \(\lambda\lambda\) 5577 feature is present in all spectra from +1 to +31d. Neither of the features are detected in the nebular-phase spectrum at +87d. This is common in nebular spectra of Ca-rich SNe as cooling via [Ca II] emission is far more efficient than via [O I] emission -- the Einstein A coefficient of [Ca II] is two orders of magnitude greater than that of [O I] \citep{jacobson2022circumstellar}. Due to this, even small quantities of calcium will cause [Ca II] emission features to dominate the spectra. 

The two [O I] features evolve almost identically with the measured velocities of each feature agreeing at a 1.5\(\sigma\) significance level. The [O I] features have a very peculiar velocity evolution compared to the [Ca II] nebular feature.  Fig. \ref{fig:velocity_ev} shows the continuum-corrected velocity evolution and FWHM evolution of these features from +1 to +31d. The FWHM measurements at +31d are estimates of the true FWHM at this phase as the two [O I] features were too blended with the continuum to be accurately modelled. The emission features were modelled as a combination of the pseudo-continuum and a Gaussian emission profile. Unlike the [Ca II] nebular feature wherein the mean velocity at +1d is \(\sim\)2,000 km s\textsuperscript{-1}, the mean velocity of [O I] \(\lambda\lambda\) 5,577 is 6,600 \(\pm\) 200 km s\textsuperscript{-1} at the same phase. Due to the blending of spectral features around this wavelength at this phase, it was not possible to isolate and determine the velocity of the [O I] \(\lambda\lambda\) 6,300 feature. In addition to this, as neither [O I] feature is detected in the +87d spectrum, it is not possible to measure the true nebular-phase velocity of [O I] in SN 2023xwi. As with [Ca II], we would expect its nebular velocity to be consistent with 0 km s\textsuperscript{-1}. The FWHM of the two [O I] features are consistent with one another and remain constant from +1d to +14d at 3,700 \(\pm\) 300 km s\textsuperscript{-1} at a significance level of 2.1\(\sigma\). The constant FWHM of the [O I] features implies there is no significant deceleration of the ejecta due to interaction with dense external material (see Section \ref{CaII}). As the [Ca II] FWHM is greater than of the [O I] feature, the emission lines up to +14d likely originate from two distinct regions \citep{dastidar2024sn}, with the [Ca II] emitting region encompassing a broader spread of velocities \citep{prentice2022oxygen}.


\section{Progenitor system}
\label{models}

The spectral features of SNe are described by two phases: the photospheric phase and the nebular phase. Due to the physical mechanisms responsible, nebular features such as [Ca II] and [O I] should not be present in photospheric spectra. The clear presence of these features from as early as +1d in SN 2023xwi indicate the existence of Ca and O that is external to the SN. To explain this pollution from heavy synthesised elements, we introduce an interpretation of an AM CVn progenitor system.

\subsection{AM CVn progenitor}
V445 Puppis -- a known galactic helium nova -- has been the subject of many investigations due to its status as a potential progenitor system of Type Ia SNe \citep{kato2018production, nyamai2021radio, banerjee2023v445}. During periods of high-accretion rates in binary systems, the accreted He-shell around a white dwarf can reach high enough densities and temperatures to ignite and undergo a He-shell flash. Such He-novae are expected to occur in binaries of a 0.75 - 1.05 M\textsubscript{\(\odot\)} CO white dwarf and a low mass H-deficient companion \citep{wong2021mass}. The companion can either be non-degenerate or, in the case of AM CVn systems, semi- / fully degenerate \citep{roelofs2007hubble}. In sufficiently compact AM CVn systems that undergo recurrent He novae, it is possible that the primary white dwarf would accrete enough He to be a Type Ia SN progenitor \citep{patat2007upper, wang2009helium}. Our investigation is therefore focused on the validity of an AM CVn system as a progenitor system of Ca-rich SNe.

\cite{woudt2009expanding} describe the nova remnant and the outburst from a He-nova in the AM CVn system V445 Puppis. They report an equatorial dust disc that obscures the nova remnant and large polar outflow velocities of 6720 \(\pm\) 650 km s\textsuperscript{-1}. The geometry and velocity of the polar lobes of this system suggest that a V445 Puppis-type system may be responsible for the spectroscopic behaviour observed in SN 2023xwi. 

\begin{figure*}
	\includegraphics[width=0.8\linewidth]{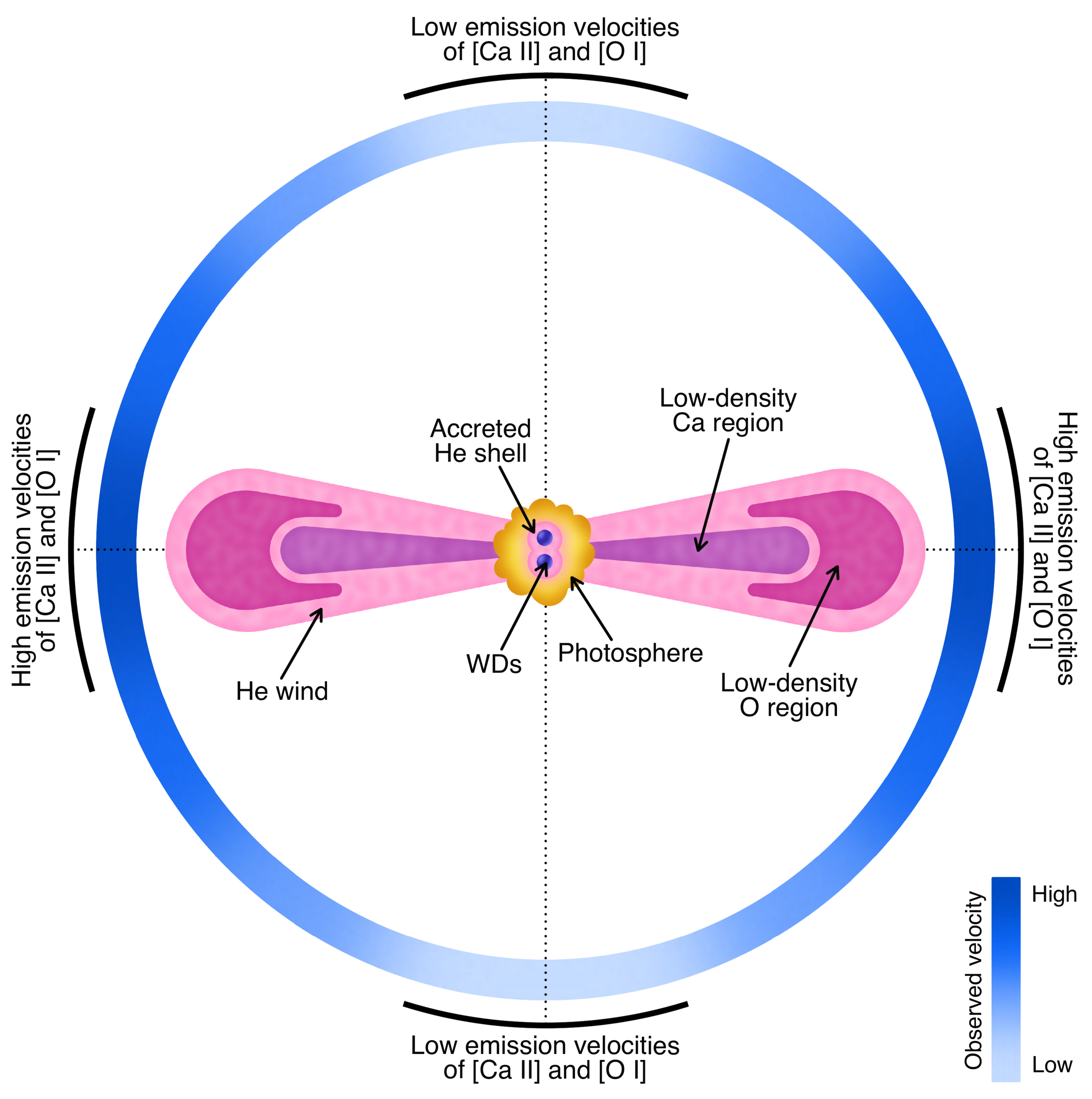}
    \caption{A 2D schematic of our proposed progenitor system of SN 2023xwi, adapted from the V445 Puppis system. Accreting binary CO white dwarfs (blue spheres) are surrounded by a shell of accreted He (light pink). After recurrent He novae, a He wind is formed in polar lobes (light pink lobes). O from the surface of the CO white dwarfs is swept up into the lobes as low-density O regions (dark pink). In more violent disruptions, Ca is swept into the lobes as low-density Ca regions (purple). Expected observable [Ca II] and [O I] emission velocities after the white dwarf explodes are included as a gradient from high velocity (dark blue) to low velocity (light blue).}
    \label{fig:puppis_type_model}
\end{figure*}

Fig. \ref{fig:puppis_type_model} shows a schematic of a V445 Puppis-type progenitor system, adapted to explain the observed behaviour of SN 2023xwi. The primary white dwarf in the double-degenerate binary system accretes He from its companion and undergoes recurrent He-novae. In each event, He is expelled and forms polar He-wind lobes, as in \cite{woudt2009expanding}. Mixing between freshly-accreted He, O from the surface of the white dwarf, and the ashes of earlier He-novae causes material from the primary white dwarf to be swept into the lobes in subsequent He-novae. This could form low-density O regions at \(\sim\) wind velocities (i.e. similar to the polar outflow velocity of V445 Puppis: 6720 \(\pm\) 650 km s\textsuperscript{-1}). In more violent He-novae, it is also possible for elements up to Ca to be synthesised and ejected into the lobes \citep{kato2018production}. As this matter is synthesised in violent He-novae, and not directly swept up by the winds, the low-density Ca regions within the lobes will have a lower velocity than the low-density O regions and the He wind. After recurrent novae, the primary white dwarf will accrete enough He from its companion to undergo a SN event \citep{patat2007upper, wang2009helium}. In the SN, these low-density Ca and O regions would initially be farther from the centre of the explosion than the ejecta, and therefore would produce observable [Ca II] and [O I] emission features that would not be `obscured' by the photosphere. As the ejecta expand beyond the reach of the lobes and the photosphere recedes, the true nebular [Ca II] and [O I] features of the SN would dominate the spectra. Such a system could therefore explain not only the presence of [Ca II] and [O I] emission in the photospheric phase, but also the observed velocity and FWHM evolution of these features in SN 2023xwi.

\subsubsection{Early [Ca II] and [O I] emission}

Cooling via [O I] emission is relatively inefficient compared to cooling via [Ca II] emission (see Section \ref{OI}). If the [O I] and [Ca II] emissions originated from the same region, the system would therefore require high quantities of low-density oxygen in this region for the emission to be discernible from the continuum. In addition to this, the FWHM evolution of the [Ca II] and [O I] features in SN 2023xwi suggest that the early [Ca II] and [O I] emission originates from two distinct regions (see Section \ref{OI}). This is explained in a V445 Puppis-type system as the regions of low density O and Ca extend to different radii. As the two emission regions in such a system are essentially separate, this would allow [O I] emission from lower abundances of O without it being dominated by [Ca II] emission. \cite{kato2018production} found that violent He-novae synthesise only small amounts of Ca, with fractional abundances of the total ejecta mass X(Ca) = 10\textsuperscript{-5} - 10\textsuperscript{-4}. As line cooling via [Ca II] emission is a very efficient mechanism, only trace amounts of Ca are required to produce the observed spectra of SN 2023xwi.

\subsubsection{[Ca II] and [O I] velocity evolution}

The photospheric velocities of [Ca II] and [O I] in SN 2023xwi impose constraints on a V445 Puppis-type progenitor system. The mean [O I] \(\lambda\lambda\) 5577 velocity of SN 2023xwi at +1d from \emph{c}-band peak is 6,600 \(\pm\) 200 km  s\textsuperscript{-1}. The polar outflow velocities in V445 Puppis were determined to be 6720 \(\pm\) 650 km s\textsuperscript{-1}, and therefore it is feasible that a similar system could result in the outflow velocities required for SN 2023xwi \citep{woudt2009expanding}. As O is less massive than Ca, the velocity of the low-density O region will be greater than the velocity of the low-density Ca region if synthesised by the same, or similar, novae. The observed mean [Ca II] velocity (1600 \(\pm\) 100 km s\textsuperscript{-1}) is significantly lower than that of [O I], and therefore is feasible in a V445 Puppis-type system. At +31d, the observed mean [O I] velocity is 3,700 \(\pm\) 200 km/s -- much greater than the [Ca II] velocity at this same phase of 300 \(\pm\) 100 km/s. In a Puppis-type system, low-density O regions are expected to extend further from the binary WD system than the low-density Ca regions (see Fig. \ref{fig:puppis_type_model}). We would therefore expect for the high velocity [O I] features to exist longer than the high velocity [Ca II] features as it would take more time for the low-density O regions to be overtaken by the SN ejecta. As the [O I] emission feature was not detected at +87d, its mean nebular velocity could not be investigated. It is expected that the associated velocity would be consistent with 0 km s\textsuperscript{-1}, as is the case for [Ca II], but this cannot be confirmed without direct observations. If non-zero nebular-phase velocities are observed in other Ca-rich SNe, this would cast doubt on the validity of the progenitor system. 

The majority of O in the observable [O I] region will, however, have been swept into the lobes by the He wind in previous He-novae, and not ejected into the lobes in the same events that synthesise Ca. The more-violent novae that produce Ca will also contain O in their ejecta, creating a mixed region of O, Ca, and other synthesised elements. Any emission from O mixed into the Ca regions would be dominated by [Ca II] emission, and thus strong [O I] emission from this mixed region would not be observed.

The outflow velocity of V445 Puppis was determined along the pole. As the system is intrinsically non-spherical, if such a system is observed off-axis, there will be a considerable discrepancy between the observed velocity and the true velocity of the emitting regions due to line-of-sight projection effects. These velocity constraints on the system are therefore dependent upon the angle of the observer relative to the system. In addition to this, as V445 Puppis is the only observed recurrent He nova to date, it is plausible that similar systems could have polar outflows with higher velocities than the V445 Puppis system. It is also likely that the geometry of similar systems could vary significantly from V445 Puppis (e.g. the lobes could be far wider or narrower). Without further observations of the behaviour of the [Ca II] and [O I] features, we cannot determine the observation angle of SN 2023xwi relative to the proposed progenitor system. 

\subsubsection{Future work}

\begin{figure}
	\includegraphics[width=\columnwidth]{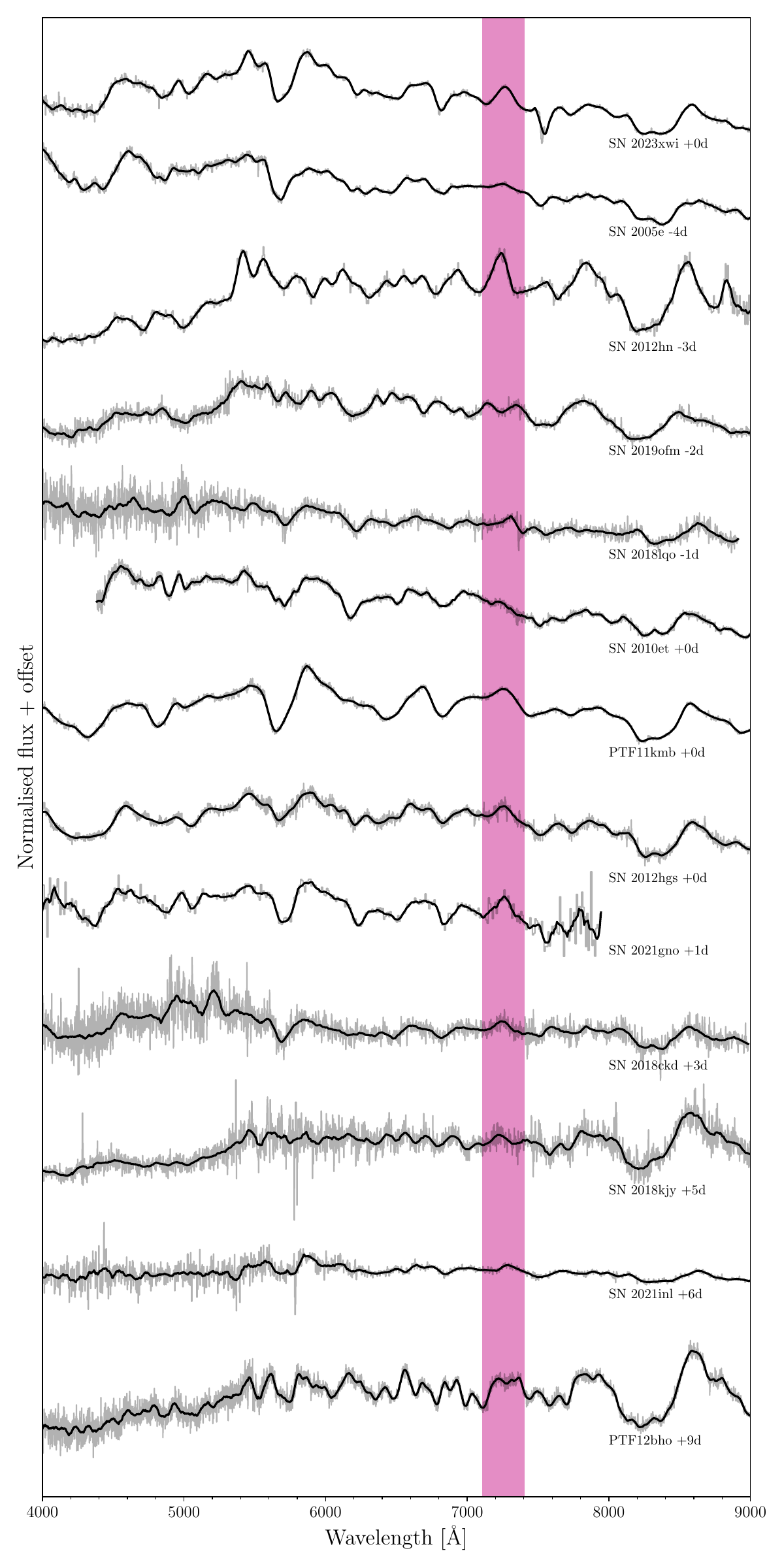}
    \caption{Near-peak rest-frame spectra of SN 2023xwi and a number of known Ca-rich SNe (labelled). All spectra  shown were taken within 9d of r-band maximum. The location of the expected [Ca II] feature in each spectrum is highlighted in dark pink.}
    \label{fig:peak_spectra}
\end{figure}

\cite{de2020zwicky} present spectroscopic and nebular spectra for a number of known Ca-rich SNe. Despite it not being a focus of their investigation, we noticed that many of the objects displayed the same early [Ca II] feature as SN 2023xwi. Fig. \ref{fig:peak_spectra} shows the spectra near peak - within 9d from r-band maximum - of a number of known Ca-rich SNe. The presence of the [Ca II] feature in many of these objects indicates pre-existing low-density Ca has polluted the environment of these SNe. A V445 Puppis-type system may, therefore, be an appropriate progenitor system for all Ca-rich SNe, supporting the argument that Ca-rich SNe are a heterogeneous group \citep{prentice2020rise, ertini2023sn}. To investigate the validity of this progenitor system, further observations of the velocity evolution of the [Ca II] and [O I] features are required. 

Assuming the progenitor system is correct, two distinct regions in velocity space produce [Ca II] emission -- the low density Ca region and the SN explosion itself. The combined emission from these two regions could therefore create a complex profile in the [Ca II] feature in the nebular phase. If the photosphere has receded before the SN ejecta overtakes the low-density lobes, we would expect to observe double-emission in the [Ca II] and [O I] features - one component from the external low-density regions, and one component from the core of the explosion. Conversely, if the photosphere has not receded before it overtakes the lobes, we would not expect to observe emission features from either of the low-density regions; in this scenario, the photosphere would effectively `obscure' the lobes. We do not observe either of these transition features in SN 2023xwi. This is likely due to the gaps in spectroscopic data between +6d and +14d, and between +31d and +87d. Direct observation of these predicted transitional features would support the validity of a V445 Puppis-type progenitor system. 

In addition to the velocity evolution of early [Ca II] and [O I] features, the mean velocity at - or near - peak would provide constraints for such a progenitor system. We expect the highest observed [Ca II] and [O I] velocities to correspond to observations directly down the polar lobes, and the lowest velocities to be observed perpendicular to these lobes (see Fig. \ref{fig:puppis_type_model}). Determining the velocities at peak of the full sample of Ca-rich SNe would place better constraints on the peak velocities we would expect from these systems. While the number of known Ca-rich SNe is small, the distribution of velocities at (or close to) peak brightness would constrain the geometry of these systems. If the velocity distribution is peaked at observation angles along the polar lobes -- i.e. if very few Ca-rich SNe show high velocity [Ca II] and [O I] features -- the lobes are likely to be very narrow. If the velocity distribution is instead more gradual -- i.e. if most [Ca II] and [O I] features have varied, but distinctly non-zero, velocities -- the lobes will be broader.

If the angle of observation relative to the progenitor system can be determined, the `true' velocity of the low-density regions within the lobes would provide valuable insight into the validity of this model. For example, a V445 Puppis-type system would not be viable if the velocity of the low-density Ca and O regions within the lobes is greater than the measured photospheric velocity of the SN. For this to be the case, a He-nova would have to accelerate the same species to a higher velocity than a far more energetic SN -- this is implausible. Direct observation of this would therefore disprove this progenitor system. 

These open questions regarding the nature of the velocity evolution of these [Ca II] and [O I] features, as well as their velocities at peak, remain the subjects of future work.


\section{Summary}
\label{summary}

In this paper, we have presented optical photometric and spectroscopic observations of a newly-discovered Ca-rich SN: SN 2023xwi. Its absolute \emph{r}-band magnitude at peak of -15.8 \(\pm\) 0.1 mag, rise time of 15.9 \(\pm\) 0.2 days, photospheric velocity of 10,900 \(\pm\) 100 km s\textsuperscript{-1}, and strong nebular-phase [Ca II] emission features confirm its status as a member of this class of exotic SNe. Below we summarise the key observations and peculiarities of SN 2023xwi, and propose a new progenitor system for Ca-rich SNe.

\begin{itemize}
    \item SN 2023xwi was detected 5 days post-explosion by the MASTER network and was spectroscopically classified as a Ca-rich SN at 8 days post-explosion by Padova-Asiago. It has a large offset of 8.91 kpc from its host galaxy PSO J146.8867+71.2435.
    \item The spectroscopic behaviour of SN 2023xwi displays the earliest and most prominent forbidden [Ca II] feature in photospheric spectra of Ca-rich SNe at +1d relative to \emph{c}-band maximum. From +6d relative to \emph{c}-band peak there is also definitive presence of forbidden [O I] emission. These features are typically associated with the nebular phase of SN observations, and are not expected to be observed in Ca-rich SNe until \(\sim\)30d post-peak \citep{kasliwal2012calcium, de2020zwicky}. The velocity evolution of these features negates the possibility of an external `blob' of ejecta being responsible for these nebular features. There is a possibility that we have misidentified our photospheric phase emission lines. Spectroscopic modelling by \cite{dessart2015one} show a strong Ti II feature and a Sc II feature in the region of the [Ca II] and [O I] features, respectively. These models, however, do not paint a complete picture of SN 2023xwi due to line blanketing effects that cause these models to be redder than our object, and other literature examples of Ca-rich SNe. Our analysis continued under the assumption that we correctly identified the features in SN 2023xwi as strong [Ca II] and [O I].
    \item The photometric behaviour exhibits an early flux excess as seen in many other objects of this class. It displayed a peak magnitude of -15.8 \(\pm\) 0.1 mag at MJD 60276.1, and a peak bolometric luminosity of 2.2 \(\pm\) 0.3 \(\times\) 10\textsuperscript{41} erg s\textsuperscript{-1}. This value aligns with other objects of this class, confirming it is not an uncharacteristic luminosity which causes the peculiar spectroscopic behaviour of SN 2023xwi.
    \item Using the photometric and spectroscopic observations, we determined the total ejecta mass to be M\textsubscript{ej} = 0.9 \(\pm\) 0.2 M\textsubscript{\(\odot\)}, with M\textsubscript{Ni} = 0.02 \(\pm\) 0.01 M\textsubscript{\(\odot\)} of synthesised \textsuperscript{56}Ni and M\textsubscript{Ca} = \(9 \times 10^{-4} \) M\textsubscript{\(\odot\)} of synthesised Ca II. All of these values are consistent with other Ca-rich SNe, ruling out the possibility that SN 2023xwi's unique spectroscopic behaviour is caused by abnormal amounts of Ca or Ni.
    \item We present a SN progenitor embedded in an environment polluted by a recurrent He-nova AM CVn system, based on V445 Puppis, that can explain the observed spectroscopic behaviour of SN 2023xwi and all other Ca-rich SNe. This model explains the presence of [Ca II] and [O I] nebular emission lines in the photospheric spectra of SN 2023xwi, as well as the observed velocity evolution of these emission features. To test the validity of a V445 Puppis-type system, spectroscopic analysis into the evolution of the forbidden [Ca II] and [O I] features should be conducted across a full sample of Ca-rich SNe. These will be investigated in later works.
\end{itemize}


\section*{Acknowledgements}
We thank Christian Knigge for useful scientific discussions. C.F. and A.P. acknowledge support from a Royal Society International Exchanges grant (IES\textbackslash\,R\textbackslash\,3223075). M.P. and J.L acknowledge support from a UK Research and Innovation Fellowship (MR/T020784/1). L.G. acknowledges financial support from AGAUR, CSIC, MCIN and AEI 10.13039/501100011033 under projects PID2023-151307NB-I00, PIE 20215AT016, CEX2020-001058-M, and 2021-SGR-01270. TLK acknowledges support via an Research Council of Finland grant (340613; P.I. R. Kotak), and from the UK Science and Technology Facilities Council (STFC, grant number ST/T506503/1).

Based in part on observations made with the Nordic Optical Telescope, owned in collaboration by the University of Turku and Aarhus University, and operated jointly by Aarhus University, the University of Turku and the University of Oslo, representing Denmark, Finland and Norway, the University of Iceland and Stockholm University at the Observatorio del Roque de los Muchachos, La Palma, Spain, of the Instituto de Astroﬁsica de Canarias. The NOT data presented here were obtained with ALFOSC, which is provided by the Instituto de Astroﬁsica de Andalucia (IAA) under a joint agreement with the University of Copenhagen and NOT.

The Gravitational-wave Optical Transient Observer (GOTO) project acknowledges the support of the Monash-Warwick Alliance; University of Warwick; Monash University; University of Sheffield; University of Leicester; Armagh Observatory \& Planetarium; the National Astronomical Research Institute of Thailand (NARIT); Instituto de Astrofísica de Canarias (IAC); University of Portsmouth; University of Turku. We acknowledge support from the Science and Technology Facilities Council (STFC, grant numbers ST/T007184/1, ST/T003103/1, ST/T000406/1, ST/X001121/1 and ST/Z000165/1).

This work makes use of observations from the Las Cumbres Observatory Global Telescope network. Time on the Las Cumbres Observatory 1m network was awarded via OPTICON (proposal 23B030)

This work makes use of observations from the Las Cumbres Observatory global telescope network.

Based on observations made with the Liverpool Telescope operated on the island of La Palma by Liverpool John Moores University in the Spanish Observatorio del Roque de los Muchachos of the Instituto de Astrofisica de Canarias with financial support from the UK Science and Technology Facilities Council.

This work includes data from the Asteroid Terrestrial-impact Last Alert System (ATLAS) project. ATLAS is primarily funded to search for near earth asteroids through NASA grants NN12AR55G, 80NSSC18K0284, and 80NSSC18K1575; byproducts of the NEO search include images and catalogs from the survey area. The ATLAS science products have been made possible through the contributions of the University of Hawaii Institute for Astronomy, the Queen's University Belfast, the Space Telescope Science Institute, and the South African Astronomical Observatory. 

The Pan-STARRS1 Surveys (PS1) and the PS1 public science archive have been made possible through contributions by the Institute for Astronomy, the University of Hawaii, the Pan-STARRS Project Office, the Max-Planck Society and its participating institutes, the Max Planck Institute for Astronomy, Heidelberg and the Max Planck Institute for Extraterrestrial Physics, Garching, The Johns Hopkins University, Durham University, the University of Edinburgh, the Queen's University Belfast, the Harvard-Smithsonian Center for Astrophysics, the Las Cumbres Observatory Global Telescope Network Incorporated, the National Central University of Taiwan, the Space Telescope Science Institute, the National Aeronautics and Space Administration under Grant No. NNX08AR22G issued through the Planetary Science Division of the NASA Science Mission Directorate, the National Science Foundation Grant No. AST-1238877, the University of Maryland, Eotvos Lorand University (ELTE), the Los Alamos National Laboratory, and the Gordon and Betty Moore Foundation.

\section*{Data Availability}
The data underlying this article will be shared on reasonable request to the corresponding author.




\bibliographystyle{mnras}
\bibliography{xwi} 

\begin{thebibliography}{}
\makeatletter
\relax
\def\mn@urlcharsother{\let\do\@makeother \do\$\do\&\do\#\do\^\do\_\do\%\do\~}
\def\mn@doi{\begingroup\mn@urlcharsother \@ifnextchar [ {\mn@doi@} {\mn@doi@[]}}
\def\mn@doi@[#1]#2{\def\@tempa{#1}\ifx\@tempa\@empty \href {http://dx.doi.org/#2} {doi:#2}\else \href {http://dx.doi.org/#2} {#1}\fi \endgroup}
\def\mn@eprint#1#2{\mn@eprint@#1:#2::\@nil}
\def\mn@eprint@arXiv#1{\href {http://arxiv.org/abs/#1} {{\tt arXiv:#1}}}
\def\mn@eprint@dblp#1{\href {http://dblp.uni-trier.de/rec/bibtex/#1.xml} {dblp:#1}}
\def\mn@eprint@#1:#2:#3:#4\@nil{\def\@tempa {#1}\def\@tempb {#2}\def\@tempc {#3}\ifx \@tempc \@empty \let \@tempc \@tempb \let \@tempb \@tempa \fi \ifx \@tempb \@empty \def\@tempb {arXiv}\fi \@ifundefined {mn@eprint@\@tempb}{\@tempb:\@tempc}{\expandafter \expandafter \csname mn@eprint@\@tempb\endcsname \expandafter{\@tempc}}}

\bibitem[\protect\citeauthoryear{Akaike}{Akaike}{1974}]{akaike1974new}
Akaike H.,  1974, IEEE transactions on automatic control, 19, 716

\bibitem[\protect\citeauthoryear{Arendse et~al.,}{Arendse et~al.}{2023}]{arendse2023detecting}
Arendse N.,  et~al., 2023, arXiv preprint arXiv:2312.04621

\bibitem[\protect\citeauthoryear{Arnett}{Arnett}{1982}]{arnett1982type}
Arnett W.~D.,  1982, Astrophysical Journal, Part 1, vol. 253, Feb. 15, 1982, p. 785-797., 253, 785

\bibitem[\protect\citeauthoryear{Banerjee, Evans, Woodward, Starrfield, Su, Ashok  \& Wagner}{Banerjee et~al.}{2023}]{banerjee2023v445}
Banerjee D.,  Evans A.,  Woodward C.,  Starrfield S.,  Su K.,  Ashok N.,   Wagner R.,  2023, The Astrophysical Journal Letters, 952, L26

\bibitem[\protect\citeauthoryear{Bazin et~al.,}{Bazin et~al.}{2009}]{bazin2009core}
Bazin G.,  et~al., 2009, Astronomy \& Astrophysics, 499, 653

\bibitem[\protect\citeauthoryear{Bora, Vink{\'o}  \& K{\"o}nyves-T{\'o}th}{Bora et~al.}{2022}]{bora2022initial}
Bora Z.,  Vink{\'o} J.,   K{\"o}nyves-T{\'o}th R.,  2022, Publications of the Astronomical Society of the Pacific, 134, 054201

\bibitem[\protect\citeauthoryear{Bradley et~al.,}{Bradley et~al.}{2024}]{larry_bradley_2024_10967176}
Bradley L.,  et~al., 2024, astropy/photutils: 1.12.0, \mn@doi{10.5281/zenodo.10967176}, \url {https://doi.org/10.5281/zenodo.10967176}

\bibitem[\protect\citeauthoryear{Branch}{Branch}{1998}]{branch1998type}
Branch D.,  1998, Annual Review of Astronomy and Astrophysics, 36, 17

\bibitem[\protect\citeauthoryear{Cardelli, Clayton  \& Mathis}{Cardelli et~al.}{1989}]{cardelli1989relationship}
Cardelli J.~A.,  Clayton G.~C.,   Mathis J.~S.,  1989, Astrophysical Journal, Part 1 (ISSN 0004-637X), vol. 345, Oct. 1, 1989, p. 245-256., 345, 245

\bibitem[\protect\citeauthoryear{Chambers et~al.,}{Chambers et~al.}{2016}]{chambers2016pan}
Chambers K.~C.,  et~al., 2016, arXiv preprint arXiv:1612.05560

\bibitem[\protect\citeauthoryear{Dastidar et~al.,}{Dastidar et~al.}{2024}]{dastidar2024sn}
Dastidar R.,  et~al., 2024, Astronomy \& Astrophysics, 685, A44

\bibitem[\protect\citeauthoryear{De et~al.,}{De et~al.}{2018}]{de2018iptf}
De K.,  et~al., 2018, The Astrophysical Journal, 866, 72

\bibitem[\protect\citeauthoryear{De et~al.,}{De et~al.}{2020}]{de2020zwicky}
De K.,  et~al., 2020, The Astrophysical Journal, 905, 58

\bibitem[\protect\citeauthoryear{Dessart \& Hillier}{Dessart \& Hillier}{2015}]{dessart2015one}
Dessart L.,  Hillier D.~J.,  2015, Monthly Notices of the Royal Astronomical Society, 447, 1370

\bibitem[\protect\citeauthoryear{Dimitriadis et~al.,}{Dimitriadis et~al.}{2023}]{dimitriadis2023sn}
Dimitriadis G.,  et~al., 2023, Monthly Notices of the Royal Astronomical Society, 521, 1162

\bibitem[\protect\citeauthoryear{Dyer et~al.,}{Dyer et~al.}{2024}]{dyer2024gravitational}
Dyer M.~J.,  et~al., 2024, in Ground-based and Airborne Telescopes X. pp 858--865

\bibitem[\protect\citeauthoryear{Ertini et~al.,}{Ertini et~al.}{2023}]{ertini2023sn}
Ertini K.,  et~al., 2023, Monthly Notices of the Royal Astronomical Society, 526, 279

\bibitem[\protect\citeauthoryear{Fisher, Branch, Nugent  \& Baron}{Fisher et~al.}{1997}]{fisher1997evidence}
Fisher A.,  Branch D.,  Nugent P.,   Baron E.,  1997, The Astrophysical Journal, 481, L89

\bibitem[\protect\citeauthoryear{Frohmaier, Sullivan, Maguire  \& Nugent}{Frohmaier et~al.}{2018}]{frohmaier2018volumetric}
Frohmaier C.,  Sullivan M.,  Maguire K.,   Nugent P.,  2018, The Astrophysical Journal, 858, 50

\bibitem[\protect\citeauthoryear{Gress et~al.,}{Gress et~al.}{2023}]{gress2023master}
Gress O.,  et~al., 2023, Transient Name Server Discovery Report, 2023, 1

\bibitem[\protect\citeauthoryear{Jacobson-Gal{\'a}n et~al.,}{Jacobson-Gal{\'a}n et~al.}{2020a}]{jacobson2020hnk}
Jacobson-Gal{\'a}n W.~V.,  et~al., 2020a, The Astrophysical Journal, 896, 165

\bibitem[\protect\citeauthoryear{Jacobson-Gal{\'a}n et~al.,}{Jacobson-Gal{\'a}n et~al.}{2020b}]{jacobson2020sn}
Jacobson-Gal{\'a}n W.~V.,  et~al., 2020b, The Astrophysical Journal, 898, 166

\bibitem[\protect\citeauthoryear{Jacobson-Gal{\'a}n et~al.,}{Jacobson-Gal{\'a}n et~al.}{2022}]{jacobson2022circumstellar}
Jacobson-Gal{\'a}n W.,  et~al., 2022, The Astrophysical Journal, 932, 58

\bibitem[\protect\citeauthoryear{Jeffery \& Branch}{Jeffery \& Branch}{1990}]{jeffery1990analysis}
Jeffery D.~J.,  Branch D.,  1990, Supernovae, 6th Jerusalem Winter School for Theoretical Physics at Jerusalem, Israel, 28 December 1988--05 January 1989, p.~149

\bibitem[\protect\citeauthoryear{Kasliwal et~al.,}{Kasliwal et~al.}{2012}]{kasliwal2012calcium}
Kasliwal M.~M.,  et~al., 2012, The Astrophysical Journal, 755, 161

\bibitem[\protect\citeauthoryear{Kato, Saio  \& Hachisu}{Kato et~al.}{2018}]{kato2018production}
Kato M.,  Saio H.,   Hachisu I.,  2018, The Astrophysical Journal, 863, 125

\bibitem[\protect\citeauthoryear{Khatami \& Kasen}{Khatami \& Kasen}{2019}]{khatami2019physics}
Khatami D.~K.,  Kasen D.~N.,  2019, The Astrophysical Journal, 878, 56

\bibitem[\protect\citeauthoryear{Kumar et~al.,}{Kumar et~al.}{2021}]{kumar2021sn}
Kumar A.,  et~al., 2021, Monthly Notices of the Royal Astronomical Society, 502, 1678

\bibitem[\protect\citeauthoryear{Lunnan et~al.,}{Lunnan et~al.}{2017}]{lunnan2017two}
Lunnan R.,  et~al., 2017, The Astrophysical Journal, 836, 60

\bibitem[\protect\citeauthoryear{Lyman, Levan, Church, Davies  \& Tanvir}{Lyman et~al.}{2014}]{lyman2014progenitors}
Lyman J.,  Levan A.~J.,  Church R.,  Davies M.~B.,   Tanvir N.,  2014, Monthly Notices of the Royal Astronomical Society, 444, 2157

\bibitem[\protect\citeauthoryear{Maeda et~al.,}{Maeda et~al.}{2010}]{maeda2010asymmetric}
Maeda K.,  et~al., 2010, Nature, 466, 82

\bibitem[\protect\citeauthoryear{Masci et~al.,}{Masci et~al.}{2018}]{masci2018zwicky}
Masci F.~J.,  et~al., 2018, Publications of the Astronomical Society of the Pacific, 131, 018003

\bibitem[\protect\citeauthoryear{Milisavljevic et~al.,}{Milisavljevic et~al.}{2017}]{milisavljevic2017iptf15eqv}
Milisavljevic D.,  et~al., 2017, The Astrophysical Journal, 846, 50

\bibitem[\protect\citeauthoryear{Mulchaey, Kasliwal  \& Kollmeier}{Mulchaey et~al.}{2013}]{mulchaey2013calcium}
Mulchaey J.~S.,  Kasliwal M.~M.,   Kollmeier J.~A.,  2013, The Astrophysical Journal Letters, 780, L34

\bibitem[\protect\citeauthoryear{Newville, Stensitzki, Allen, Rawlik, Ingargiola  \& Nelson}{Newville et~al.}{2016}]{newville2016lmfit}
Newville M.,  Stensitzki T.,  Allen D.~B.,  Rawlik M.,  Ingargiola A.,   Nelson A.,  2016, Astrophysics Source Code Library, pp ascl--1606

\bibitem[\protect\citeauthoryear{Nicholl}{Nicholl}{2018}]{nicholl2018superbol}
Nicholl M.,  2018, Research Notes of the AAS, 2, 230

\bibitem[\protect\citeauthoryear{Nyamai, Chomiuk, Ribeiro, Woudt, Strader  \& Sokolovsky}{Nyamai et~al.}{2021}]{nyamai2021radio}
Nyamai M.,  Chomiuk L.,  Ribeiro V.,  Woudt P.,  Strader J.,   Sokolovsky K.,  2021, Monthly Notices of the Royal Astronomical Society, 501, 1394

\bibitem[\protect\citeauthoryear{Orfanidis}{Orfanidis}{1995}]{orfanidis1995introduction}
Orfanidis S.~J.,  1995, Introduction to signal processing.
Prentice-Hall, Inc.

\bibitem[\protect\citeauthoryear{Parrent, Branch  \& Jeffery}{Parrent et~al.}{2010}]{parrent2010synow}
Parrent J.,  Branch D.,   Jeffery D.,  2010, Astrophysics Source Code Library, pp ascl--1010

\bibitem[\protect\citeauthoryear{Patat et~al.,}{Patat et~al.}{2007}]{patat2007upper}
Patat F.,  et~al., 2007, Astronomy \& Astrophysics, 474, 931

\bibitem[\protect\citeauthoryear{Perets et~al.,}{Perets et~al.}{2010}]{perets2010faint}
Perets H.~B.,  et~al., 2010, Nature, 465, 322

\bibitem[\protect\citeauthoryear{Piro \& Nakar}{Piro \& Nakar}{2013}]{piro2013can}
Piro A.~L.,  Nakar E.,  2013, The Astrophysical Journal, 769, 67

\bibitem[\protect\citeauthoryear{Polin, Nugent  \& Kasen}{Polin et~al.}{2019}]{polin2019observational}
Polin A.,  Nugent P.,   Kasen D.,  2019, The Astrophysical Journal, 873, 84

\bibitem[\protect\citeauthoryear{Polin, Nugent  \& Kasen}{Polin et~al.}{2021}]{polin2021nebular}
Polin A.,  Nugent P.,   Kasen D.,  2021, The Astrophysical Journal, 906, 65

\bibitem[\protect\citeauthoryear{Prentice et~al.,}{Prentice et~al.}{2020}]{prentice2020rise}
Prentice S.,  et~al., 2020, Astronomy \& Astrophysics, 635, A186

\bibitem[\protect\citeauthoryear{Prentice, Maguire, Siebenaler  \& Jerkstrand}{Prentice et~al.}{2022}]{prentice2022oxygen}
Prentice S.,  Maguire K.,  Siebenaler L.,   Jerkstrand A.,  2022, Monthly Notices of the Royal Astronomical Society, 514, 5686

\bibitem[\protect\citeauthoryear{Price-Whelan et~al.,}{Price-Whelan et~al.}{2022}]{price2022astropy}
Price-Whelan A.~M.,  et~al., 2022, The Astrophysical Journal, 935, 167

\bibitem[\protect\citeauthoryear{Roelofs, Groot, Benedict, McArthur, Steeghs, Morales-Rueda, Marsh  \& Nelemans}{Roelofs et~al.}{2007}]{roelofs2007hubble}
Roelofs G.~H.,  Groot P.,  Benedict G.,  McArthur B.,  Steeghs D.,  Morales-Rueda L.,  Marsh T.,   Nelemans G.,  2007, The Astrophysical Journal, 666, 1174

\bibitem[\protect\citeauthoryear{Schlafly \& Finkbeiner}{Schlafly \& Finkbeiner}{2011}]{schlafly2011measuring}
Schlafly E.~F.,  Finkbeiner D.~P.,  2011, The Astrophysical Journal, 737, 103

\bibitem[\protect\citeauthoryear{Schlegel, Finkbeiner  \& Davis}{Schlegel et~al.}{1998}]{schlegel1998maps}
Schlegel D.~J.,  Finkbeiner D.~P.,   Davis M.,  1998, The Astrophysical Journal, 500, 525

\bibitem[\protect\citeauthoryear{Shen, Quataert  \& Pakmor}{Shen et~al.}{2019}]{shen2019progenitors}
Shen K.~J.,  Quataert E.,   Pakmor R.,  2019, The Astrophysical Journal, 887, 180

\bibitem[\protect\citeauthoryear{Shingles et~al.,}{Shingles et~al.}{2021}]{shingles2021release}
Shingles L.,  et~al., 2021, Transient Name Server AstroNote, 7, 1

\bibitem[\protect\citeauthoryear{Steeghs et~al.,}{Steeghs et~al.}{2022}]{steeghs2022gravitational}
Steeghs D.,  et~al., 2022, Monthly Notices of the Royal Astronomical Society, 511, 2405

\bibitem[\protect\citeauthoryear{Stone}{Stone}{1979}]{stone1979comments}
Stone M.,  1979, Journal of the Royal Statistical Society. Series B (Methodological), pp 276--278

\bibitem[\protect\citeauthoryear{Taubenberger et~al.,}{Taubenberger et~al.}{2019}]{taubenberger2019sn}
Taubenberger S.,  et~al., 2019, Monthly Notices of the Royal Astronomical Society, 488, 5473

\bibitem[\protect\citeauthoryear{Thomas, Nugent  \& Meza}{Thomas et~al.}{2011}]{thomas2011synapps}
Thomas R.,  Nugent P.,   Meza J.,  2011, Publications of the Astronomical Society of the Pacific, 123, 237

\bibitem[\protect\citeauthoryear{Tonry et~al.,}{Tonry et~al.}{2018}]{tonry2018atlas}
Tonry J.,  et~al., 2018, Publications of the Astronomical Society of the Pacific, 130, 064505

\bibitem[\protect\citeauthoryear{Wang, Meng, Chen  \& Han}{Wang et~al.}{2009}]{wang2009helium}
Wang B.,  Meng X.,  Chen X.,   Han Z.,  2009, Monthly Notices of the Royal Astronomical Society, 395, 847

\bibitem[\protect\citeauthoryear{Wong \& Bildsten}{Wong \& Bildsten}{2021}]{wong2021mass}
Wong T. L.~S.,  Bildsten L.,  2021, The Astrophysical Journal, 923, 125

\bibitem[\protect\citeauthoryear{Woudt et~al.,}{Woudt et~al.}{2009}]{woudt2009expanding}
Woudt P.,  et~al., 2009, The Astrophysical Journal, 706, 738

\bibitem[\protect\citeauthoryear{Zenati, Perets, Dessart, Jacobson-Gal{\'a}n, Toonen  \& Rest}{Zenati et~al.}{2023}]{zenati2023origins}
Zenati Y.,  Perets H.~B.,  Dessart L.,  Jacobson-Gal{\'a}n W.~V.,  Toonen S.,   Rest A.,  2023, The Astrophysical Journal, 944, 22

\makeatother
\end{thebibliography}



\bsp	
\label{lastpage}
\end{document}